\documentclass[twocolumn,showpacs,preprintnumbers,amsmath,amssymb]{revtex4}
\usepackage{graphicx}

\begin{document}

\title{Localization of massive fermions on the baby-skyrmion branes in 6-dimensions}
\author{Yuta Kodama}
\author{Kento Kokubu}
\author{Nobuyuki Sawado}
\email{sawado@ph.noda.tus.ac.jp}
\affiliation{Department of Physics, Tokyo University of Science, 
Noda, Chiba 278-8510, Japan}

\date{\today}

\begin{abstract}
We construct brane solutions in 6-dimensional Einstein-Skyrme systems. 
A class of baby skyrmion solutions realizes warped compactification of the extra 
dimensions and gravity localization on the brane for negative bulk cosmological constant. 
Coupling of the fermions with the brane skyrmions lead to the 
brane localized fermions. 
In terms of the level crossing picture, 
emergence of the massive localized modes are observed.  
Nonlinear nature of the skyrmions brings richer information 
for the fermions level structure. It comprises doubly degenerate lowest plus 
single excited modes. The three generation of the fundamental fermions is associated 
with this distinctive structure. 
The mass hierarchy of quarks or leptons is appeared in terms of a slightly deformed 
baby-skyrmions with topological charge three. 
\end{abstract} 
\maketitle 

\section{Introduction}
Theories with extradimensions have been expected to solve the hierarchy 
problem and cosmological constant problem. Experimentally unobserved 
extradimensions indicate that the standard model particles and forces 
are confined to a 3-brane~\cite{ArkaniHamed:1998rs,ArkaniHamed:1998nn,Randall:1999ee,Randall:1999vf}. 
Intensive study has been performed for the Randall-Sundram (RS) brane model in 5 space-time 
dimensions~\cite{Randall:1999ee,Randall:1999vf}. In this framework, the exponential 
warp factor in the metric can generate a large hierarchy of scales. 
This model, however, requires unstable negative tension branes  
and the fine-tuning between brane tensions and bulk cosmological constant.
 
There is hope that higher dimensional brane models more than five 
could evade those problems appeared in 5-dimensions. 
In fact brane theories in 6-dimensions show a very distinct feature towards 
the fine-tuning and negative tension brane problems. 
In Refs.~\cite{Carroll:2003db,Navarro:2003vw}, it was shown that the brane tension merely 
produces deficit angles in the bulk and hence it can take an arbitrary value without 
affecting the brane geometry. The model is based on the spontaneous compactification 
by the bulk magnetic flux. If the compactification manifold is a sphere, two branes 
have to be introduced with equal tensions. If it is a disk, no second 3-brane is 
needed. But still the fine-tuning between magnetic flux and the bulk cosmological constant  
cannot be avoided although non-static solutions could be free of any fine-tuning~\cite{Kanti:2001vb}.  

Alternatively to the flux compactification in 6-dimensions, the nonlinear sigma model has been 
used for compactifications of the extra space 
dimensions~\cite{GellMann:1984sj,Kehagias:2004fb,RandjbarDaemi:2004ni,Lee:2004vn}. 
As in the flux compactification, no second 3-brane is needed if the parameters 
in the sigma model and bulk space-time are tuned.   

Warped compactifications are also possible in 6 space-time dimensions 
in the model of topological objects such as defects and solitons.  
In this context strings~\cite{Cohen:1999ia,Gregory:1999gv,Gherghetta:2000qi,Giovannini:2001hh,Peter:2003zg} 
were investigated, showing that they can realize localization of gravity.  
Interestingly, if the brane is modeled in such a field theory language, 
the fine-tuning between bulk and brane parameters required in the case 
of delta-like branes turns to a tuning of the model parameters~\cite{Peter:2003zg}. 

The Skyrme model is known to possess soliton solutions called baby skyrmions 
in 2-dimensional space~\cite{Piette:1994ug,Kudryavtsev:1996er}. In this paper we therefore consider 
the warped compactification of the 2-dimensional extra space by the baby skyrmions.  
We find that in the 6-dimensional Einstein-Skyrme systems, static solutions which realize warped 
compactification exist for negative bulk cosmological constant. Since the solution 
is regular except at the conical singularity, it has only single 3-brane. 
Thus no fine-tuning between brane tensions is required. The Skyrme model possess a 
rich class of stable multi soliton solutions. We find various brane solutions by such 
multi-solitons.
 
It should be noted that general considerations in the 6-dimensional brane model with bulk scalar fields 
suggest that the mechanism of regular warped compactification with single 
positive tension brane is not possible~\cite{Chen:2000at}. However, the model under consideration 
is restricted to the bulk scalar field depending only on the radial coordinate in the extra 
space. The scalar field in the Skyrme model depend not only the radial coordinate but 
also the angular coordinate to exhibit nontrivial topological structure, which 
makes possible to realize regular warped compactification.  

Study of localization of fermions and gauge fields on topological defects have been extensively 
studied with co-dimension one \cite{Kehagias:2000au,Melfo:2006hh,Ringeval:2001cq,Koley:2008dh,Hosotani:2006qp} 
and two \cite{RandjbarDaemi:2000cr,Libanov:2000uf,Neronov:2001qv,
RandjbarDaemi:2003qd,Parameswaran:2006db,Aguilar:2006sz,Zhao:2007aw,Guo:2008ia}. 
Many years ago, particle localization on a domain wall in higher dimensional
space time was already addressed \cite{Rubakov:1983bz,Akama:1982jy}. 
The authors suggested the possibility of localized massless fermions 
on the 1-dimensional kink background in 4+1 space-time with Yukawa-type coupling manner.
Later, localization of chiral fermions on RS scenario was discussed in Ref.\cite{Kehagias:2000au}. 
Analysis for the massive fermionic modes was done by Ringeval {\it et.al.}, in Ref.\cite{Ringeval:2001cq}.
For co-dimension two, the localization on higher dimensional generalizations of the RS model
was studied within the coupling of real scalar fields \cite{RandjbarDaemi:2000cr}.
Many studies have been followed and most of them are based on the 
Abelian Higgs, or Higgs mediated models with the chiral fermions. 

Problem of fermion mass hierarchy has been discussed in many articles
\cite{ArkaniHamed:1999dc,RandjbarDaemi:2000cr,Dvali:2000ha,Libanov:2000uf,Neronov:2001qv,Hung:2001hw,
Aguilar:2006sz,Hosotani:2006qp,Guo:2008ia} 
within the different mechanisms. They are based on the Yukawa coupling 
of the fermions and the Higgs (scalar) field. In Ref.\cite{Dvali:2000ha}, 
the authors set up multiple branes and considered localization of fermions on different branes 
in terms of Yukawa couplings to the Higgs field. In Ref.\cite{Neronov:2001qv}, 
the fermions have quantum numbers of the rotational momenta which are origin of 
the generation of fermions. 
The authors of Ref.\cite{Aguilar:2006sz} deal with this problem with somewhat different 
approach. Conical singularity of the background branes and the orbital angular momentum 
of the fermions around the branes are the key role for the generation.  
In Ref.\cite{Libanov:2000uf} hierarchy between the fermionic generations are explained 
in terms of multi-winding number solutions of the complex scalar fields. 
They observed three chiral fermionic zero modes on a topological defect with winding 
number three and finite masses appear the mixing of these zero modes. 
Although any brane localization mechanism is absent in their discussion, 
the idea is promising. 
In Ref.\cite{Hung:2001hw}, the authors have taken into account more realistic 
standard model charges. 
Ref.\cite{Guo:2008ia} is about discussion of the fermion families from two layer warped
6-dimensions. The authors obtained Kaluza-Klein particles in 5 dimensions at first; 
those are finally regarded as light, standard model fermions in 4 dimensions. 

Starting point of our approach is conventional and has somewhat similarity 
to Ref.\cite{Libanov:2000uf}.
We shall consider the localization of the fermions on the baby skyrmion branes with 
topological charge three. The localized modes of fermions are confirmed through 
the analysis of spectral flow of the one particle state \cite{Kahana:1984be}. 
According to the Index theorem a nonzero topological charge implies the 
zero modes of the Dirac operator \cite{Atiyah:1980jh}. 
The zero crossing modes are found to be the localized fermions on the brane. 
So the generation of the fermions is defined in terms of the topological 
charge of the skyrmions with a special quantum number called grandspin $K_3$. 
There are different profiles of the zero crossing behavior for different $K_3$, 
and it is the origin of the finite mass in our point of view.
Nonlinear nature of the skyrmion fields has richer information than the case of Abelian string; 
the level comprises lowest doubly degenerate as well as single excited modes, 
which can partially explains generation puzzle of the fermions.  
In order to manifest more realistic mass structure, 
breaking of the rotational symmetry and the shape deformation of 
the background skyrmions is taken into account. 

The crucial difference of our approach from the other attempts is the representation of 
the fermions. It is based on our knowledge that even in the first generation of the fermions, 
they have small but finite masses. It means that the fermions are not pure chiral eigenstates.   
Therefore, in this article we employ the standard representation of the higher dimensional 
gamma matrices instead of the chiral one. 

This paper is organized as follows. In the next section we describe the 
Einstein-Skyrme system in 6-dimensions and derives the coupled equations for 
the Skyrme and gravitational fields. We derive a class of multi-winding number 
solutions. Some typical numerical brane solutions are shown. 
Formulation of the fermions in higher dimensional curved space-time is discussed in Sec.III. 
Coupling of the fermions and the skyrmions is introduced in this section.  
Conclusion and discussion are given in Sec.IV.

\section{\label{sec:2} Construction of the baby-skyrmion branes}
\subsection{\label{subsec:level11}Model}
We introduce a model of the 6-dimensional Einstein-Skyrme system with a bulk cosmological constant 
coupled to fermions \cite{Kodama:2008zza}. The action comprises 
\begin{eqnarray}
	S=S_{\rm gravity}+S_{\rm brane}+S_{\rm fermion}\label{action}\,.
\end{eqnarray}
Here $S_{\rm gravity}$ is the 6-dimensional Einstein-Hilbert action
\begin{eqnarray} 
	S_{\rm gravity}&=&\int d^{6}x \sqrt{-g}\left[\frac{1}{2\kappa^2}R-\Lambda_{b}\right]\,.
	\label{grav_action} 
\end{eqnarray}
In the parameter $\kappa^2=1/M_{6}^{4}$, $M_{6}$ is the 6-dimensional 
Planck mass, denoted the fundamental gravity scale, and 
$\Lambda_{b}$ is the bulk cosmological constant.

For $S_{\rm brane}$ we use the action of baby-Skyrme model~\cite{Piette:1994ug,Kudryavtsev:1996er}
\begin{eqnarray}
S_{\rm brane}&=&\int d^{6}x {\cal L}_{\rm brane}
\end{eqnarray}
with
\begin{eqnarray}
	&&{\cal L}_{\rm brane}=\sqrt{-g}\Bigl[\frac{F^2}{2}\partial_{M}
	{\vec \phi}\cdot \partial^{M}{\vec \phi}
	+\frac{1}{4e^2}\bigl(\partial_{M}{\vec \phi}\times\partial_{N}
	{\vec \phi}\bigr)^{2} \nonumber \\
	&&\hspace{5cm}+\mu^2 (1+{\vec n}\cdot{\vec \phi})\Bigr]\,,
	\label{skyrme_action}
\end{eqnarray}
where $M,N$ run over $0,\cdots ,5$ and  ${\vec n}=(0,0,1)$.
 ${\vec \phi}=(\phi^{1},\phi^{2},\phi^{3})$ denotes a triplet of scalar real fields with 
the constraint ${\vec \phi}\cdot{\vec \phi}=1$.
The constants $F,e,\mu$ are the Skyrme model parameters with the dimension of
 $(energy)^{2}$, $(energy)^{-1}$, $(energy)^{3}$, 
respectively. The first term in Eq.(\ref{skyrme_action}) is nothing but a nonlinear $\sigma-$model.   
The second term is the analogue of the Skyrme fourth order term in the standard Skyrme model
(the Skyrme model in 3+1 dimensions) 
which works as a stabilizer for obtaining the soliton solution. The last term is referred to
as a potential term which guarantee the stability of a baby-skyrmion. 

The solutions of the 
model would be characterized by following topological charge in curved space-time
\begin{eqnarray}
Q=\frac{1}{4\pi}\int d^2x{\vec \phi}\cdot (\nabla_1{\vec \phi}\times \nabla_2{\vec \phi})
\label{windingnumber}
\end{eqnarray}
where $\nabla_\mu$ means the space-time covariant derivative. 
Let us assume that the matter Skyrme fields depend only on the 
extra coordinates and impose the hedgehog ansatz 
\begin{eqnarray}
      {\vec \phi}=(\sin f(r)\cos n\theta,\, \sin f(r)\sin n\theta,\, \cos f(r))\,. 
      \label{hedgehog}
\end{eqnarray}
The function $f(r)$ which is often called as the profile function, 
has following boundary condition
\begin{eqnarray}
f(0)=-(m-1)\pi,~~\lim_{r\to \infty}f(r)=\pi
\end{eqnarray}
where $(m,n)$ is arbitrary integer. This ansatz ensures the topological charge
\begin{eqnarray}
Q=n(1-(-1)^m)/2.
\label{windingnumber2}
\end{eqnarray}

We consider the maximally symmetric metric with vanishing 4D cosmological constant, 
\begin{eqnarray}
	ds^{2}=B^{2}(r)\eta_{\mu\nu}dx^{\mu}dx^{\nu} + dr^{2}
	+ C^{2}(r)d\theta^{2} \label{metric}
\end{eqnarray}
where $\eta_{\mu\nu}$ is the Minkowski metric with the signature 
$(-,+,+,+)$ in our convention and $0\le r < \infty$ and
 $0\le \theta \le 2\pi$. 
This ansatz has been proved to realize warped compactification 
of the extra dimension in models where branes are represented 
by global defects~\cite{Olasagasti:2000gx}. 

$S_{\rm fermion}$ is the action for fermions coupled with the skyrmions and the gravity; 
we shall describe it in Sec.\ref{sec:level3}.

\begin{figure}[t]
\includegraphics[height=7cm, width=9cm]{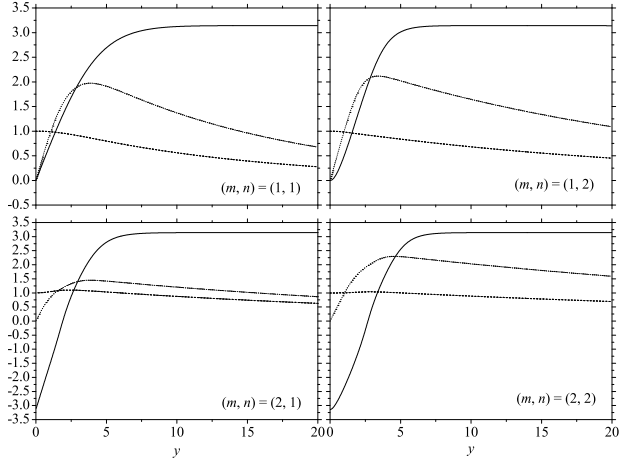}
\caption{\label{profilef} Typical results of the profile functions $f$ (straight line), 
the warp metrices $B,\tilde{C}$ (dashed,dotted line,respectively), 
as a function of $y$. }
\end{figure}

The general forms of the coupled system of Einstein equations and the 
equation of motion of the Skyrme model are 
\begin{eqnarray}
&&G_{MN}=\kappa^2 (-\Lambda_{b} g_{MN} +T_{MN})\,, \\
&&\frac{1}{\sqrt{-g}}\partial_N\Bigl(\sqrt{-g}F^2{\vec \phi}\times \partial^N{\vec \phi} \nonumber \\
&&+\sqrt{-g}\frac{1}{e^2}\partial_M{\vec \phi}\bigl(\partial^M{\vec \phi}\cdot({\vec \phi}\times \partial^N{\vec \phi})\bigr)\Bigr)
+\mu^2{\vec \phi}\times{\vec n}=0\,,\nonumber \\
\end{eqnarray}
where the stress-energy tensor $T_{MN}$ is given by
\begin{eqnarray}
&&T_{MN}=-2\frac{\delta {\cal L}_{\rm brane}}{\delta g^{MN}}+g_{MN}{\cal L}_{\rm brane} \nonumber \\
&&=F^2\partial_{M}{\vec \phi}\cdot \partial_{N}{\vec \phi}
	+\frac{1}{e^2}g^{AB}\bigl(\partial_{A}{\vec \phi}\times\partial_{M}{\vec \phi}\bigr)
	\cdot\bigl(\partial_{B}{\vec \phi}\times\partial_{N}{\vec \phi}\bigr)\nonumber \\
	&&+g_{MN}{\cal L}_{\rm brane}\,.
	\label{stress_tensor}
\end{eqnarray}
Inserting Eq.~(\ref{hedgehog}) into Eq.~(\ref{skyrme_action}), one obtains 
the Lagrangian 
\begin{eqnarray}
	{\cal L}_{\rm brane}=-B^4{\tilde C}F^4e^2\left[uf'^{2}+\frac{n^2\sin^{2}f}{{\tilde C^{2}}}
	+2{\tilde \mu}(1+\cos f)\right] \nonumber \\
	\label{skyrme-lag}
\end{eqnarray}
where we have introduced the dimensionless quantities 
\begin{eqnarray}
	\tilde{x}_\mu=eFx_\mu,~y=eFr,~{\tilde C}=eFC,~
	{\tilde \mu}=\frac{1}{eF^2}\mu \label{}
\end{eqnarray}
and 
\begin{eqnarray}
	u=1+\frac{n^2\sin^{2}f}{{\tilde C}^{2}}\,. \label{}
\end{eqnarray}
The prime denotes derivative with respect to the radial component $y$ of the two extra space. 
The Skyrme field equation is thus
\begin{eqnarray}
	&&f''+\left(\frac{4B'}{B}+\frac{{\tilde C}'}{{\tilde C}}+\frac{u'}{u}
	\right)f' \nonumber \\
	&&-\frac{1}{2u}\left[\frac{n^2\sin 2f}{{\tilde C}^{2}}
	(1+f'^{2})+2{\tilde \mu}^2\sin f\right]=0 \label{skyrme}
\end{eqnarray}
where 
\begin{eqnarray}
	\frac{u'}{u}=\frac{n^2}{{\tilde C}^{2}+n^2\sin^{2}f}\left[f'\sin 2f
	-2\frac{{\tilde C}'}{{\tilde C}}\sin^{2}f\right]\,. \label{}
\end{eqnarray}
Within this ansatz, the components of the stress-energy tensor
 (\ref{stress_tensor}) becomes
\begin{eqnarray}
&&T_{\mu\nu}=-F^4e^2B^2\eta_{\mu\nu}\tau_0(y)\,,~~\nonumber \\
&&~~~~\tau_0(y)\equiv \frac{u}{2}f'^2+\frac{n^2\sin^2f}{2\tilde{C}^2}+\tilde{\mu}^2(1+\cos f)
\label{estensor0}\\
&&T_{rr}=-F^4e^2\tau_r(y)\,,~~\nonumber \\
&&~~~~\tau_r(y)\equiv -\frac{u}{2}f'^2+\frac{n^2\sin^2f}{2\tilde{C}^2}+\tilde{\mu}^2(1+\cos f)
\label{estensorr}\\
&&T_{\theta\theta}=-F^4\tilde{C}^2\tau_\theta(y)\,,\nonumber \\
&&~~~~\tau_\theta(y)\equiv \frac{\hat{u}}{2}f'^2-\frac{n^2\sin^2f}{2\hat{C}^2}+\tilde{\mu}^2(1+\cos f)
\label{estensort}
\end{eqnarray}
where 
\begin{eqnarray}
	\hat{u}=1-\frac{n^2\sin^{2}f}{{\tilde C}^{2}}\,. \label{}
\end{eqnarray}
The Einstein equations with bulk cosmological constant are written 
down in the following form
\begin{eqnarray}
	&& 3{\hat b}'+6{\hat b}^{2}+3{\hat b}\,{\hat c}+{\hat c}'+{\hat c}^2
	=-\alpha({\tilde \Lambda}_{b}+\tau_0(y)) \label{einstein1}\\ 
	&& 6{\hat b}^{2}+4{\hat b}\,{\hat c}=-\alpha
	({\tilde \Lambda}_{b}+\tau_r(y)) \label{einstein2}\\
	&& 4{\hat b}'+10{\hat b}^{2}=-\alpha({\tilde \Lambda}_{b}
	+\tau_\theta(y))\label{einstein3}
\end{eqnarray}
where $\alpha=\kappa^2 F^2$ is a dimensionless coupling constant and 
${\tilde \Lambda}_{b}=\Lambda_{b} /e^2F^{4}$ is a dimensionless bulk cosmological constant. 
Also, we introduce ${\hat b}\equiv B'/B,{\hat c}\equiv C'/C$ for convenience. 

\begin{figure}[t]
\includegraphics[height=7.0cm, width=9cm]{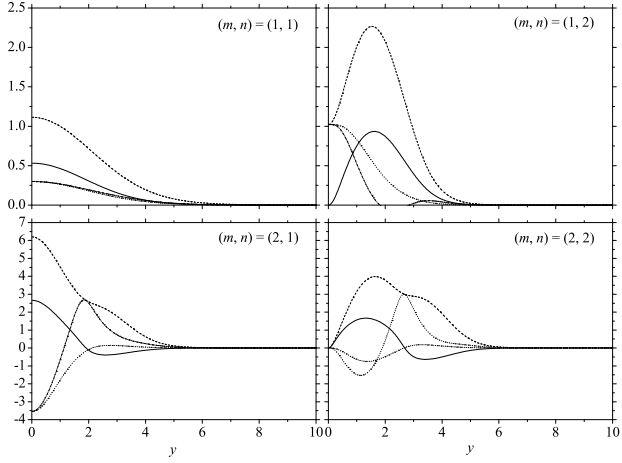}
\caption{\label{stensor} Typical results of the stress-energy tensors $\tau_0,\tau_r,\tau_\theta$ 
(dashed,dotted,and dot-dashed line, respectively) and the topological charge
density $q$ (straight line).}
\end{figure}

\subsection{Boundary conditions}
At infinity, all components of the energy-momentum tensor vanishes and the Einstein 
equations~(\ref{einstein1})-(\ref{einstein3}) are then reduced to 
\begin{eqnarray}
	&&3{\hat b}'+6{\hat b}^{2}+3{\hat b}\,{\hat c}+{\hat c}'+{\hat c}^2
	=-\alpha {\tilde \Lambda}_{b} \\
	&&6{\hat b}^{2}+4{\hat b}\,{\hat c}=-\alpha 
	{\tilde \Lambda}_{b} \\
	&&4{\hat b}'+10{\hat b}^{2}=-\alpha {\tilde \Lambda}_{b}\,.\label{}
\end{eqnarray}
The general solution has been obtained in Refs.~\cite{Giovannini:2001hh,Peter:2003zg} which 
is given by
\begin{eqnarray}
	{\hat b}=p\,\frac{Ae^{\frac{5}{2}py}-e^{-\frac{5}{2}py}}
	{Ae^{\frac{5}{2}py}+e^{-\frac{5}{2}py}}\;,\;\;\;\;\;
	{\hat c}=\frac{5p^{2}}{2{\hat b}}
	-\frac{3}{2}{\hat b} \label{}
\end{eqnarray}
where $A$ is an arbitrary constant and 
\begin{eqnarray}
	p=\sqrt{\frac{-\alpha {\tilde \Lambda}_{b}}{10}}\,. \label{}
\end{eqnarray}
Since we are interested in regular solutions with warped compactification of the 
extra-space, the functions $B$ and ${\tilde C}$ must converge at infinity. 
This is achieved only when ${\tilde \Lambda}_{b}<0$ and $A=0$ with the solution
\begin{eqnarray}
	B \rightarrow \epsilon_{1}\,e^{-py}
	\;,\;\;\;\;
	{\tilde C} \rightarrow \epsilon_{2}\,e^{-py} \label{bound-inf1}
\end{eqnarray}
where $\epsilon_{1}$ and $\epsilon_{2}$ are arbitrary constants.
Then, the asymptotic form of the metric which realizes warped compactification 
is given by 
\begin{eqnarray}
	ds_{\infty}^{2}&=&\epsilon_{1}e^{-2\sqrt{\frac{-\alpha {\tilde \Lambda}_{b}}{10}}y}
	\eta_{\mu\nu}dx^{\mu}dx^{\nu} \nonumber \\
	&+&dy^{2}+\epsilon_{2}e^{-2\sqrt{\frac{-\alpha {\tilde \Lambda}_{b}}{10}}y}
	d\theta^{2}\,. 
	\label{asym-met}
\end{eqnarray}

\begin{figure}[t]
\includegraphics[height=7.0cm, width=9cm]{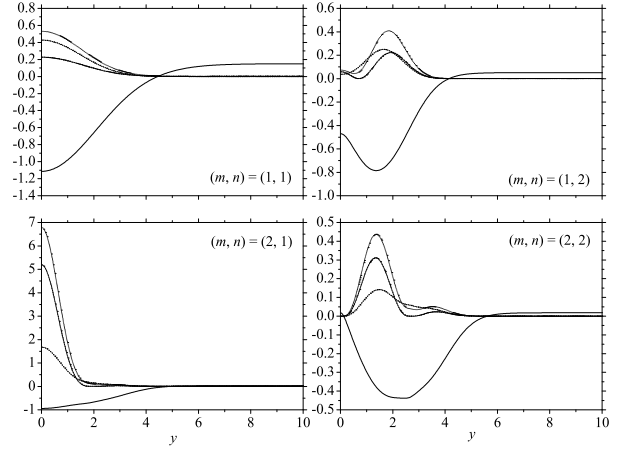}
\caption{\label{curvature} Typical results of the curvature invariants $R$, $R_{AB}R^{AB}$, $R_{ABCD}R^{ABCD}$
and $C_{ABCD}C^{ABCD}$ (straight,dashed,dotted and dot-dashed line, respectively).}
\end{figure}

The 4-dimensional reduced Planck mass $M_{pl}$ is derived by the 
coefficient of the 4 dimensional Ricci scalar, which can be calculated 
inserting the metric~(\ref{metric}) into the action~(\ref{grav_action}), 
\begin{eqnarray*}
	&&\frac{M_{pl}^{2}}{2}\int d^{4}x \sqrt{-g^{(4)}}R^{(4)}
	=\frac{M_{6}^{4}}{2}\int d^{6}x \sqrt{-g}B^{-2}(r)R^{(4)} \\
	&&=\frac{M_{6}^{4}}{2}\int d^{4}x \sqrt{-g^{(4)}}R^{(4)}
	\int dr d\theta B^{2}(r)C(r) \\
	&&= \frac{2\pi M_{6}^{4}}{2}\int dr \, B^{2}(r)C(r)
	\int d^{4}x \sqrt{-g^{(4)}}R^{(4)}  \label{}
\end{eqnarray*}
where the superscript $(4)$ represents a tensor defined on the 4 dimensional 
submanifold.  
Thus, we find the relation between $M_{pl}$ and $M_{6}$ as 
\begin{eqnarray}
	M_{pl}^{2}=2\pi M_{6}^{4}\int_{0}^{\infty}dr \,B^{2}(r) C(r) \,. 
\end{eqnarray}
The requirement of gravity localization is equivalent to the finiteness 
of the 4-dimensional Planck mass. For the asymptotic 
solution~(\ref{asym-met}), the localization is attained. 

Let us consider the asymptotic solutions for skyrmions.
we can write 
\begin{eqnarray}
f(y)=\bar{f}+\delta f(y)\,,
\end{eqnarray} 
where for $y\gg 1$, $\bar{f}\sim 0$.
The linearized field equations are given by
\begin{eqnarray}
	\delta f''-5p\delta f' -{\tilde \mu}\delta f=0\,. \label{}
\end{eqnarray} 
Assuming that $f$ falls off exponentially, one obtains for $y\gg 1$
\begin{eqnarray}
	\delta f(y) \rightarrow f_{c}e^{-qy}~~~~{\rm with}~~~~ 
	q=\frac{\sqrt{25p^{2}+4{\tilde \mu}}-5p}{2}\;\;\; 
	\label{bound-inf2}
\end{eqnarray}
where $f_{c}$ is an arbitrary constant. 

Following regularity of the geometry at the center of the defect
is imposed  
\begin{eqnarray}
B'(0)=0,~~C(0)=0,~~C'(0)=1
\end{eqnarray}
and we can arbitrarily fix $B(0)=1$.
Boundary conditions for the warp factors and the profile function at the origin are 
determined by expanding them around the origin. For the different topological sectors, 
the first few terms are schematically written down as  
\begin{eqnarray}
&&f(y)=-(m-1)\pi+f^{(n)}(0)y^n+O(y^{n+1}) \label{expf}\\
&&b(y)={\cal B}y+O(y^3) \label{expb}\\
&&\tilde{C}(y)=y+{\cal C}y^3+O(y^5) \label{expc}
\end{eqnarray}
where
\begin{eqnarray}
&&(m,n)=(1,1) \nonumber \\
&&~~~~~~{\cal B}=-\frac{\alpha}{4}\Bigl(\tilde{\Lambda}_b+2\tilde{\mu}-\frac{1}{2}f'(0)^4\Bigr)\,,\nonumber \\
&&~~~~~~{\cal C}=\frac{\alpha}{12}\Bigl(\tilde{\Lambda}_b+2\tilde{\mu}-2f'(0)^2 
-\frac{5}{2}f'(0)^4\Bigr) \\
&&(m,n)=(1,2) \nonumber \\
&&~~~~~~{\cal B}=-\frac{\alpha}{4}\bigl(\tilde{\Lambda}_b+2\tilde{\mu}),~~
{\cal C}=\frac{\alpha}{12}\bigl(\tilde{\Lambda}_b+2\tilde{\mu}) \\
&&(m,n)=(2,1) \nonumber \\
&&~~~~~~{\cal B}=-\frac{\alpha}{4}\Bigl(\tilde{\Lambda}_b-\frac{1}{2}f'(0)^4\Bigr),\nonumber \\
&&~~~~~~{\cal C}=\frac{\alpha}{12}\Bigl(\tilde{\Lambda}_b
-2f'(0)^2-\frac{5}{2}f'(0)^4\Bigr) \\
&&(m,n)=(2,2) \nonumber \\
&&~~~~~~{\cal B}=-\frac{\alpha}{4}\tilde{\Lambda}_b,~~
{\cal C}=\frac{\alpha}{12}\tilde{\Lambda}_b\,.
\end{eqnarray}
Thus one finds that the only $f'(0)$ or $f''(0)$ is the free parameter vicinity of the origin.

Consider linear combinations of Eqs.(\ref{einstein1})-(\ref{einstein3}), we obtain
\begin{eqnarray}
&&{\hat b}'+4{\hat b}^2+{\hat b}\,{\hat c}=
-\frac{1}{2}\alpha \tilde{\Lambda}_b
+\frac{\alpha}{4}(\tau_r+\tau_\theta)\,, \label{einstein5} \\
&&4{\hat b}\,{\hat c}+{\hat c}'+{\hat c}^2=
-\frac{1}{2}\alpha \tilde{\Lambda}_b+\frac{\alpha}{4}(4\tau_0+\tau_r-3\tau_\theta)\,. \label{einstein6}
\end{eqnarray}
Integrating Eqs.(\ref{einstein5}),(\ref{einstein6}) from zero to $y_c$, we get 
\begin{eqnarray}
&&B^3(y_c)B'(y_c)\tilde{C}(y_c) \nonumber \\
&&=-\frac{\alpha}{2}\tilde{\Lambda}_b\int^{y_c}_0 B^4 \tilde{C} dy
-\frac{\alpha}{4}(\mu_r+\mu_\theta)\,,
\label{einstein7}\\
&&B^4(y_c)\tilde{C}'(y_c) \nonumber \\
&&=1-\frac{\alpha}{2}\tilde{\Lambda}_b\int^{y_c}_0 B^4 \tilde{C} dy
-\frac{\alpha}{4}(4\mu_0+\mu_r-\mu_\theta)\,.
\label{einstein8} 
\end{eqnarray}
(\ref{einstein7}) is the 6-dimensional analogue of the relation determining the 
Tolman mass whereas Eq.(\ref{einstein8}) is the generalization of the relation giving the 
angular deficit. 
Combining these the following relations are obtained in the $y_c\to \infty$
\begin{eqnarray}
\alpha\int^{\infty}_0 B^4 \frac{n^2\sin^2 f}{\tilde{C}}(1+f'^2)dy=1 
\end{eqnarray}
or
\begin{eqnarray}
\alpha\int^{\infty}_0 B^4\Bigl[\frac{n^2\sin^2 f}{\tilde{C}}+2\tilde{\Lambda}_b\tilde{C}
+2\tilde{\mu}\tilde{C}(1-\cos f)\Bigr]dy=1.\nonumber \\
\end{eqnarray}
These conditions are used for checking the numerical accuracy of our calculations.  

In order to study the singularity structure of the bulk solutions, 
the several curvature invariants are
computed \cite{Giovannini:2001hh}. 
The explicit form for the metric (\ref{metric}) are given in 
Ref.\cite{Giovannini:2000cq} 
and they are
\begin{eqnarray}
&&R:=20\hat{b}^2+8\hat{b}'+2\hat{c}'+2\hat{c}^2+8\hat{b}\hat{c} \nonumber \\
&&R_{AB}R^{AB}:= \nonumber \\
&&~~80\hat{b}^4+20\hat{b}'^2+2\hat{c}'^2+2\hat{c}^4+64\hat{b}'\hat{b}^2+4\hat{c}^2\hat{c}'
+28\hat{b}^2\hat{c}^2 \nonumber \\
&&~~+32\hat{b}^3\hat{c}+8\hat{b}\hat{b}'\hat{c}+8\hat{b}'\hat{c}'+8\hat{b}'\hat{c}^2+8\hat{b}^2\hat{c}'
+8\hat{b}\hat{c}\hat{c}'+8\hat{b}\hat{c}^3 \nonumber \\
&&R_{ABCD}R^{ABCD}:= \nonumber \\
&&~~4\hat{c}^4+40\hat{b}^4+16\hat{b}^2\hat{c}^2
+8\hat{c}^2\hat{c}'+4\hat{c}'^2+32\hat{b}^2\hat{b}'+16\hat{b}'^2
\nonumber \\
&&C_{ABCD}C^{ABCD}:=\frac{12}{5}(\hat{b}'-\hat{c}'+\hat{b}\hat{c}-\hat{c}^2)^2\,. \nonumber 
\end{eqnarray}

\begin{figure}[t]
	\includegraphics[height=7cm, width=9cm]{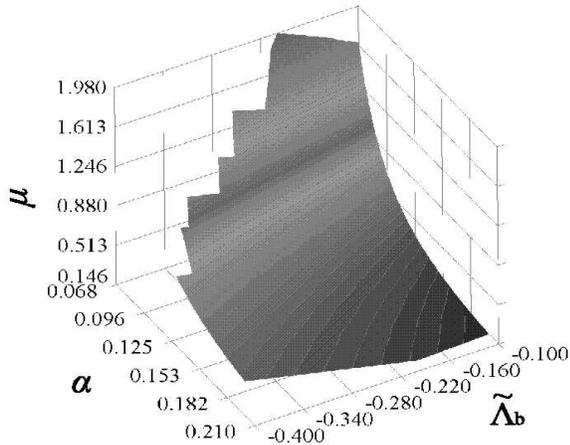}
	\caption{\label{parameter}
	Parameter space for typical solutions which exhibit gravity localization 
	around the skyrmions with $Q=3$.}
\end{figure} 

\begin{figure}[t]
	\includegraphics[height=6cm, width=8cm]{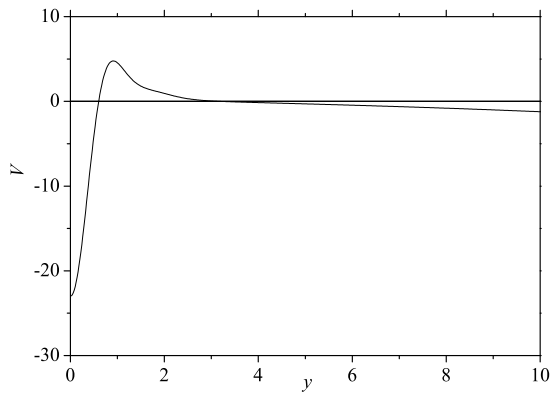}
	\caption{\label{volc}The effective potential $V_{11}$ defined by Eq.(\ref{V11})
	with the parameter,$\tilde{w}=1.1$,$l=0$ for $(m,n)=(1,1)$ is shown. The figure 
	clearly exhibits {\it Volcano} shape.}
\end{figure} 

\begin{figure}
\includegraphics[height=11cm, width=9cm]{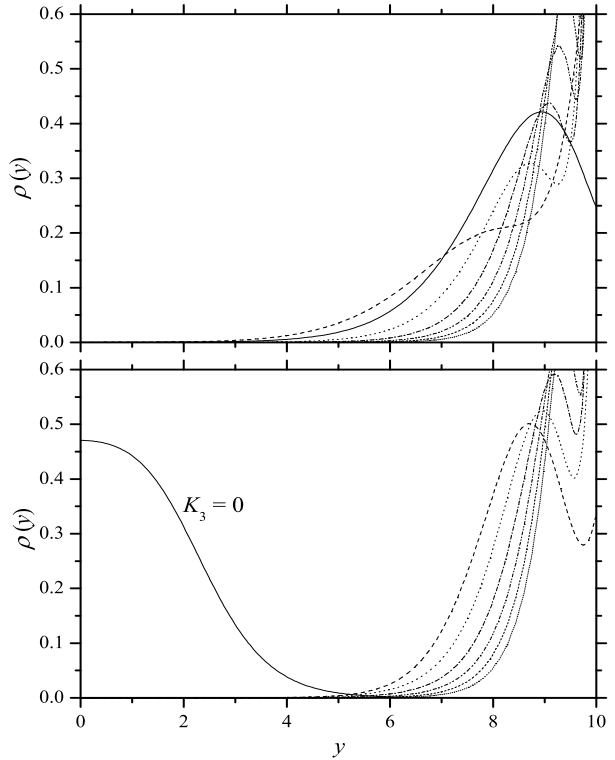}
\caption{\label{fdensity1} Fermion number density for the background skyrmion with $(m,n)=(1,1)$
with the model parameters: $\alpha=0.5,~\tilde{\Lambda}_b=-0.1,~\tilde{\mu}=0.220525915$ and 
$f'(0)=0.7292182131078184$. The results for the coupling constant $\tilde{M}=0$ (decoupled) 
and $\tilde{M}=0.86$ are plotted. 
If the fermions couple to the skyrmions, only the state $K_3=0$ is localized on the brane core.
The other states are not observed because they are strongly delocalized.}  
\end{figure}

\subsection{Numerical analysis}
The equations~(\ref{skyrme}),(\ref{einstein1})-(\ref{einstein3}) should be solved numerically since 
they are highly nonlinear. The simple technique to solve the Einstein-Skyrme 
equations is the shooting method combined with the 4th order Runge-Kutta forward 
integration~\cite{Shiiki:2005xn}. However, a unique set of boundary conditions at $y=0$ 
produces 2 distinct solutions, one of which grows exponentially and another 
decays exponentially as $y\rightarrow \infty$. This causes instability of 
solutions when the forward integration is performed. Instead, we employ a 
backward integration which is used in Ref.~\cite{Giovannini:2001hh} where the 6-dimensional 
vortex-like regular brane solutions were constructed. The backward integration 
method requires a set of boundary conditions at infinity. 
We, however, truncate and take the distance $y_{\rm max}$ far enough from the origin so that 
the Skyrme profile would fall off before it reaches $y_{\rm max}$. 
The set of boundary conditions at $y=y_{\rm max}$ produces a unique solution which 
satisfies the boundary conditions at $y=0$ and hence it is numerically stable.
We present our typical numerical results in Figs.[\ref{profilef},\ref{stensor},\ref{curvature}].
Fig.[\ref{parameter}] shows the fine-tuning surface in the model parameter space 
$(\alpha,\tilde{\mu},\tilde{\Lambda}_b)$ corresponding to the gravity localization condition.
We should stress that once the desired solutions are obtained, the parameters are no longer 
arbitrary, $i.e.$, a constraint $h(\alpha,\tilde{\mu},\tilde{\Lambda}_b)=0$ is emerged.

\section{\label{sec:level3}Fermions}
\subsection{\label{subsec:level31}Basic formalism}
The action of the fermions coupled with the Skyrme field in a Yukawa coupling manner 
can be written as
\begin{eqnarray}
S_{\rm fermion}=\int d^6x {\cal L}_{\rm fermion}
\end{eqnarray}
with
\begin{eqnarray}
{\cal L}_{\rm fermion}=\sqrt{-g}\Bigl[\bar{\Psi}(i\Gamma^AD_A-M\vec{\tau}\cdot\vec{\phi})\Psi\Bigr]\,.
\end{eqnarray}
The 6-dimensional gamma matrices $\Gamma^A$ are defined with the help the {\it vielbein} $e^A_{\hat{a}}$
and those of the flat-space $\gamma^{\hat{a}}$,{\it i.e.}, $\Gamma^A:=e^A_{\hat{a}}\gamma^{\hat{a}}$.
The covariant derivative is defined as 
\begin{eqnarray} 
D_A:=\frac{1}{2}\overleftrightarrow{\partial}_A+\frac{1}{2}\omega_A^{\hat{a}\hat{b}}\sigma_{\hat{a}\hat{b}}
\end{eqnarray}
where $\omega_A^{\hat{a}\hat{b}}:=\frac{1}{2}e^{\hat{a}B}\nabla_Ae^{\hat{b}}_B$ are the spin connection 
with generators $\sigma_{\hat{a}\hat{b}}:=\frac{1}{4}[\gamma_{\hat{a}},\gamma_{\hat{b}}]$.
The simbol $\overleftrightarrow{\partial}$ implies that $\psi\overleftrightarrow{\partial}\phi\equiv \psi\partial \phi-(\partial \psi) \phi$.
Here $A,B=0,\cdots,5$ are the 6-dimensional space-time index and $\hat{a},\hat{b}=0,\cdots,5$ corresponds to the flat
tangent 6-dimensional Minkowski space. 
The vielbein is defined through $g_{AB}=e^{\hat{a}}_Ae_{\hat{a}B}=\eta_{\hat{a}\hat{b}}e^{\hat{a}}_Ae^{\hat{b}}_B$.
We introduce the form of vielbein which was used, $e.g.$, at Ref.\cite{Zhao:2007aw}
\begin{eqnarray}
&&e^{\hat{a}}_\mu=B(r) \delta^{\hat{a}}_\mu,~~\mu=0,\cdots,3, \nonumber \\
&&e^{\hat{4}}_r=\cos \theta,~~e^{\hat{5}}_r=\sin \theta,~~\nonumber \\
&&e^{\hat{4}}_\theta=C(r)\sin\theta,~~e^{\hat{5}}_\theta=C(r)\cos\theta\,.
\end{eqnarray}
Straightforward calculation shows that the nonvanishing components of the 
corresponding spin connections are
\begin{eqnarray}
&&\omega^{\hat{\mu}\hat{4}}_\mu=\delta^{\hat{\mu}}_{\mu}B'(r)\cos\theta,~~
\omega^{\hat{\mu}\hat{5}}_\mu=\delta^{\hat{\mu}}_{\mu}B'(r)\sin\theta,~~\nonumber \\
&&\omega^{\hat{4}\hat{5}}_\theta=1-C'(r).
\end{eqnarray}
The Dirac equation is
\begin{eqnarray}
&&\Bigl[i\frac{1}{B}\delta^{\mu}_{\hat{\mu}}\gamma^{\hat{\mu}}\partial_\mu
+i(\cos\theta\gamma^{\hat{4}}+\sin\theta\gamma^{\hat{5}})
(\partial_r+\frac{2B'}{B}-\frac{1-C'}{2C}) \nonumber \\
&&-i(\sin\theta\gamma^{\hat{4}}-\cos\theta\gamma^{\hat{5}})
\frac{1}{C}\partial_\theta
-M\vec{\tau}\cdot\vec{\phi}\Bigr]\Psi=0\,.
\label{diraceq8_s}
\end{eqnarray}

\begin{figure}
\includegraphics[height=7cm, width=9cm]{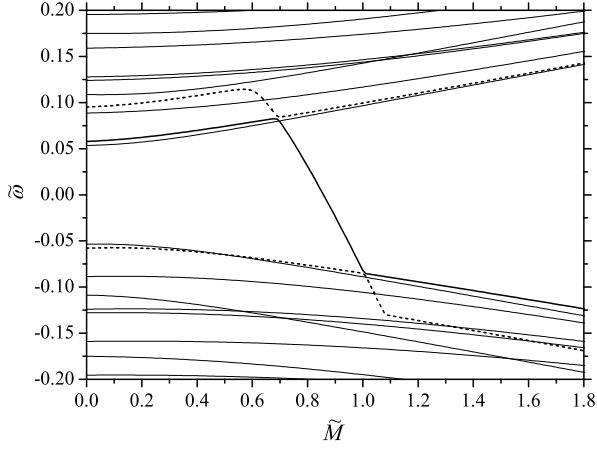}
\caption{\label{sflow1} Spectral flow of the fermion energy of $(m,n)=(1,1)$.
The zero crossing spectrum (bold line) has the quantum number $K_3=0$, 
which corresponds to the localizing zero mode. 
Exchange between a heavy (dotted line) and a light level (straight line) 
, which has been discussed in Ref.\cite{Burnier:2006za} in somewhat different context, 
is also found. }
\end{figure}

\begin{figure*}
\includegraphics[height=5.0cm, width=6.2cm]{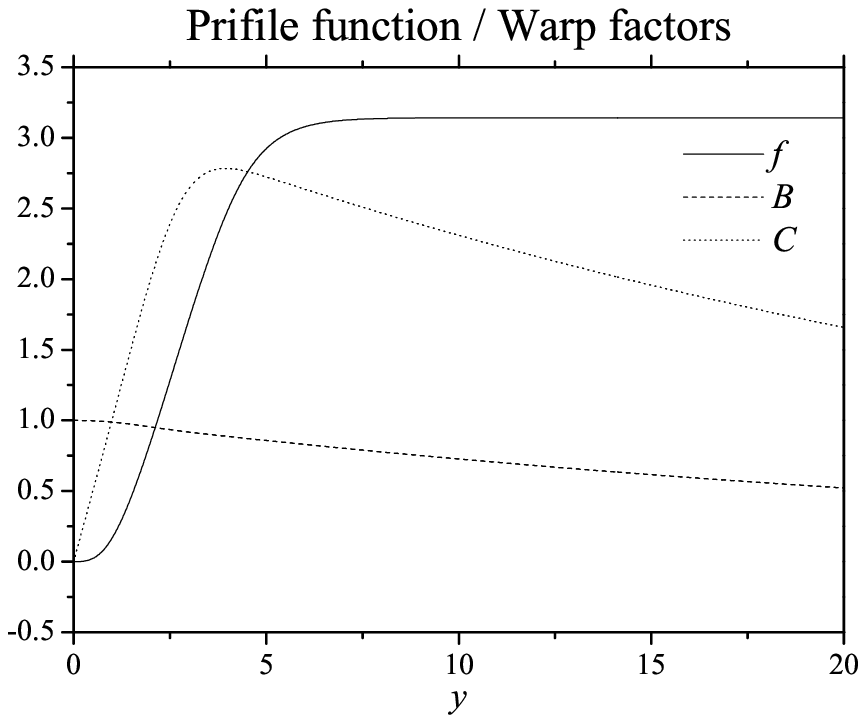}\hspace{-0.7cm}
\includegraphics[height=5.0cm, width=6.2cm]{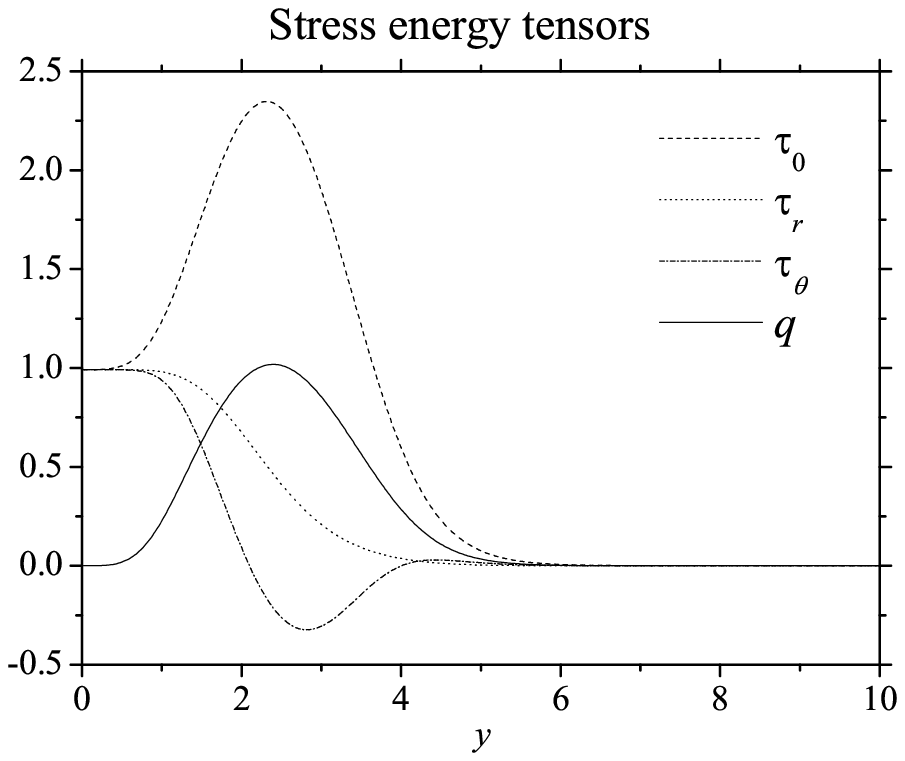}\hspace{-0.7cm}
\includegraphics[height=5.0cm, width=6.2cm]{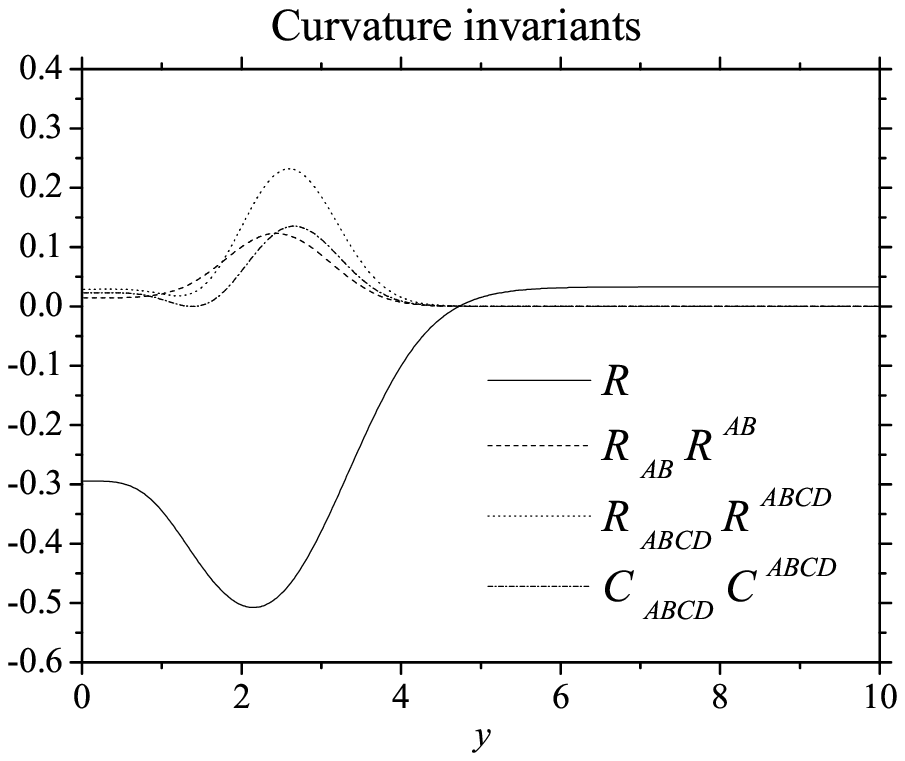}
\caption{\label{profilef3} The typical brane solution with $(m,n)=(1,3)$.
The parameters are $\alpha=0.11,~\tilde{\Lambda}_b=-0.1,~\tilde{\mu}=0.49588, 
f^{(3)}(0)=0.1913003760462098$.}
\end{figure*}

The Dirac gamma matrices should satisfy the anti-commutation relations 
$\{\gamma^{\hat{A}},\gamma^{\hat{B}}\}=2\eta^{\hat{A}\hat{B}}$
and there are the possible candidates preserving such Clifford algebra. 
In most of previous studies in the 6-dimensions, 
they are based on the localization of the chiral fermions on the Abelian vortex. 
In the chiral representation, the spinors are expanded into the right- and the 
left-handed components and the zero mode appears as a eigenstate 
of the right or the left. The massive modes emerge as their mixing states. 
Our approach, however, is somewhat different for treatment of the masses of the fermions.
Actually, even in the first generation the fermions have intrinsic, finite masses. 
So, treating the massive fermionic modes directly, we employ the standard representation of 
the higher dimensional gamma matrices instead of the chiral one. 
The eigenvalues of the Dirac hamiltonian are estimated for background brane solutions with 
large parameter space, and the zero modes appear as the zero crossing points. 
The standard representation of the gamma matrices in 6-dimensional can be defined as
\begin{eqnarray}
&&\gamma^{\hat{\mu}}:=
\left(
\begin{array}{cc}
\bar{\gamma}^{\hat{\mu}}& 0 \\
0 & -\bar{\gamma}^{\hat{\mu}} \\
\end{array}
\right),~
\bar{\gamma}^{\hat{\mu}}:= \biggl\{
\left(
\begin{array}{cc}
I_2 & 0 \\
0 & -I_2 \\
\end{array}
\right),
\left(
\begin{array}{cc}
0 & \vec{\sigma} \\
-\vec{\sigma} & 0 \\
\end{array}
\right)\biggr\} \nonumber \\
&&~~~~~~~~\gamma^{\hat{4}}:=
\left(
\begin{array}{cc}
0 & -iI_4 \\
-iI_4 & 0 \\
\end{array}
\right),
\gamma^{\hat{5}}:=
\left(
\begin{array}{cc}
0 & -I_4 \\
I_4 & 0 \\
\end{array}
\right)
\end{eqnarray}
where $I_n$ means the identity matrix of the dimension $n$.
The 6-dimensional spinor $\Psi$ can be decomposed into the 4 dimensional and 
the extra space-time components
$\Psi(x^\mu,r,\theta)=\psi(x^\mu)\otimes U(r,\theta)$.
Here the 4-dimensional part $\psi(x^\mu)$ is the solution of the corresponding 
Dirac equation on the brane
\begin{eqnarray}
i\bar{\gamma}^\mu\partial_\mu\psi=w\psi
\label{diraceq4}
\end{eqnarray}
in which the eigenvalues $w$ correspond to the 4-dimensional masses of the fermions. The 
Dirac equation in 6-dimensions is transformed as the 2-dimensional eigenproblem with the eigenvalue $w$.
The following replacement of the eigenfuncations greatly simplifies the equation of motion
~\cite{RandjbarDaemi:2003qd} 
\begin{eqnarray}
&&u(y,\theta):= \nonumber \\
&&~~~~\exp\biggl[2\ln B(y)+\frac{1}{2}\ln \tilde{C}(y)-\frac{1}{2}\int^y \frac{dy'}{\tilde{C}(y')}\biggr]
U(y,\theta)
\,. \nonumber 
\end{eqnarray}
The eigenproblem becomes
\begin{eqnarray}
H_2 u=\tilde{w}u
\label{eigenproblem8}
\end{eqnarray}
where the hamiltonian is
\begin{eqnarray}
&&\hspace{-0.5cm}H_2:=
B\left(
\begin{array}{cc}
\tilde{M}\vec{\tau}\cdot\vec{\phi} &
-e^{-i\theta}(\partial_y-\frac{i\partial_\theta}{\tilde{C}})
\\
e^{i\theta}(\partial_y+\frac{i\partial_\theta}{\tilde{C}})
&
-\tilde{M}\vec{\tau}\cdot\vec{\phi} 
\end{array}
\right)\,.
\label{hamiltonian8}
\end{eqnarray}
Here we have introduced the dimensionless coupling constant 
and the eigenvalue $\tilde{M}:=M/eF$ and $\tilde{\omega}:=\omega/eF$.
One easily confirms that $H_2$ commutes with ``grandspin''
\begin{eqnarray}
K_3:=l_3+\frac{\gamma^{\hat{6}}}{2}+n\frac{\tau^3}{2}
\end{eqnarray}
where $l_3:= -i\frac{\partial}{\partial\theta}$ is the orbital angular momentum in the extra space and 
for convenience, we introduce $\gamma^{\hat{6}}:=I\otimes \sigma^3$.
Thus the eigenstates are specified by the magnitude of the grandspin, $i.e.$, 
\begin{eqnarray}
&&K_3=0,\pm 1,\pm 2,\pm 3\cdots,~~{\rm for~odd~} n \nonumber \\
&&K_3=\pm \frac{1}{2},\pm \frac{3}{2},\pm\frac{5}{2},\cdots,~~{\rm for~even~} n\,.
\end{eqnarray}
If we consider ``time dependent'' Dirac equation
\begin{eqnarray}
i\partial_\tau \bar{u}(\tau,r,\theta)=H_2 \bar{u}(\tau,r,\theta),
\end{eqnarray} 
the equation is invariant under ``time-reversal'' transformation ${\cal T}$, $i.e.,\tau\to -\tau$,  
\begin{eqnarray}
{\cal T}H_2{\cal T}^{-1}=H_2,~~{\cal T}:=i\gamma^{\hat{5}}\otimes\tau^2 C,
\end{eqnarray}
Here $C$ is the charge conjugation operator. 
Since the hamiltonian is invariant under time reverse, 
the states of $\pm K_3$ are degenerate in energy. 

Once the desired eigenfunctions are obtained, angular averaged fermion densities on the brane can 
be estimated as follows
\begin{eqnarray}
&& \langle \rho \rangle :=\int \tilde{C} d\theta U^\dagger U
=N(y)\rho (y) \\
&&\rho (y):=\int d\theta u^\dagger(y,\theta) u(y,\theta)
\end{eqnarray}
where
\begin{eqnarray}
N(y)=\exp\biggl[-4\ln B(y)+\int^y \frac{dy'}{\tilde{C}(y')}\biggr]\,.
\end{eqnarray}
By using the asymptotics of the warp factors (\ref{bound-inf1}), 
we easily find out the behaviour $N(y)$ at zero and infinity~\cite{RandjbarDaemi:2003qd}
\begin{eqnarray}
&&N(y\to 0)\to y \nonumber \\
&&N(y\to \infty)\to \frac{1}{\epsilon_1^4}\exp(4py+\frac{1}{p\epsilon_2}e^{py})\,.  
\end{eqnarray}
At far from the core, the densities of the excited modes are more enhanced than the localized mode.

The eigenequation (\ref{eigenproblem8}) can be recast into a (Schr\"{o}dinger like) second order 
differential equation.
We assume a form of eigenfuntion
\begin{eqnarray}
	u(y,\theta):=\left(
\begin{array}{cc}
\eta_1(y)e^{il\theta} \\
\eta_2(y)e^{i(l+n)\theta} \\
\xi_1(y)e^{i(l+1)\theta} \\
\xi_2(y)e^{i(l+n+1)\theta}
\end{array}
\right) \label{eigenspinor}
\end{eqnarray}
where $l$ is an arbitrary integer. 
In order to eliminate $\xi_1$ and $\xi_2$ from Eq.(\ref{eigenproblem8}),
we write
\begin{eqnarray}
&&\xi_1=\beta(G_- D_-^l\eta_1-\tilde{M}\sin{f}D_-^{l+n}\eta_2)\\
&&\xi_2=-\beta(\tilde{M}\sin{f}D_-^l\eta_1-G_+ D_-^{l+n}\eta_2)
\end{eqnarray}
where
\begin{eqnarray}
&&\beta:=\Bigl(\frac{\tilde{\omega}^2}{B^2}-\tilde{M}^2\Bigr)^{-1} \\
&&G_\pm:=\frac{\tilde{\omega}}{B}\pm \tilde{M}\cos{f} \\
&&D_\pm^l:=\partial_y\pm \frac{l}{\tilde{C}}
\end{eqnarray}
and then, we obtain
\begin{widetext}
	\begin{eqnarray}
		&&\biggl\{\beta G_-\partial_y^2+(\delta^1G_-+\beta G'_-)\partial_y
		+\biggl(G_--\delta^{l+1}G_-\frac{l}{\tilde{C}}
		-\beta\frac{G'_-\tilde{C}-G_-\tilde{C}'}{\tilde{C}^2}l\biggr)\biggl\}\eta_1
		+\biggl\{ -\beta \tilde{M}\sin{f}\partial_y^2\nonumber\\
		&&-(\delta^{1-n}\tilde{M}\sin{f}+\beta \tilde{M}\cos{f}f')\partial_y
		-\biggl(\tilde{M}\sin{f}-\delta^{l+1}\tilde{M}\sin{f}\frac{l+n}{\tilde{C}}
		-\beta\frac{\cos{f}f'\tilde{C}-\sin{f}\tilde{C}'}{\tilde{C}^2}\tilde{M}(l+n)\biggr) \biggl\} \eta_2=0
		\label{p1}\\
		&&\biggl\{\beta \tilde{M}\sin{f}\partial_y^2
		+(\delta^{n+1}\tilde{M}\sin{f}+\beta \tilde{M}\cos{f}f')\partial_y
		+\biggl(\tilde{M}\sin{f}-\delta^{l+n+1}\tilde{M}\sin{f}\frac{l}{\tilde{C}}
		-\beta \tilde{M}\cos{f}f'\frac{l}{\tilde{C}}
		+\beta \tilde{M}\sin{f}\frac{l\tilde{C}'}{\tilde{C}^2}\biggr)\biggl\}\eta_1\nonumber\\
		&&-\biggl\{\beta G_+\partial_y^2+(\delta^{1}G_++\beta G'_+)\partial_y
		+(G_+-\delta^{l+n+1}G_+\frac{l+n}{\tilde{C}}-\beta G'_+\frac{l+n}{\tilde{C}}
		+\beta G_+\frac{(l+n)\tilde{C}'}{\tilde{C}^2})\biggl\}\eta_2=0\label{p2}
	\end{eqnarray}
	\end{widetext}
where $\delta^l:=\beta'+\frac{l}{\tilde{C}}\beta$.
Furthermore the following replacement simplifies Eqs.(\ref{p1}),(\ref{p2})
\begin{eqnarray}
&&\zeta (y):= \exp \Bigl[-\int^y \frac{{\cal R}-{\cal A}_1}{2}d\tilde{y}\Bigr] \eta_1(y),\\
&&\chi(y):=\exp \Bigl[-\int^y \frac{{\cal R}-\tilde{{\cal A}}_2}{2}d\tilde{y}\Bigr] \eta_2(y),
\end{eqnarray}
where ${\cal R}:=\frac{4B'}{B}+\frac{\tilde{C}'}{\tilde{C}}$.
We finally obtain the equation of the form
	\begin{eqnarray}
		\left(
		\begin{array}{cc}
		-\partial_y^2 -{\cal R}\partial_y+V_{11} & -{\cal A}_2\partial_y +V_{12} \\
		-\tilde{\cal A}_1\partial_y  +V_{21} & \hspace{-5mm}-\partial_y^2-{\cal R}\partial_y+V_{22} \\
		\end{array}
		\right)
		\left(
		\begin{array}{c}
		\zeta \\
		\chi \\
		\end{array}
		\right)=0\,.
		\label{eigeneq2}
	\end{eqnarray}
The explicit forms of ${\cal A}_i,\tilde{\cal{A}}_i,V_{ij}~(i,j=1,2)$ are given in appendix \ref{ap:potential}.
Here we show the typical example of the effective potential $V_{11}$ in Fig.\ref{volc}. 
The shape of the potential is like famous {\it Volcano type}. 
So we expect the existence of localized mode of the fermions. 

\begin{figure}
\includegraphics[height=7cm, width=9cm]{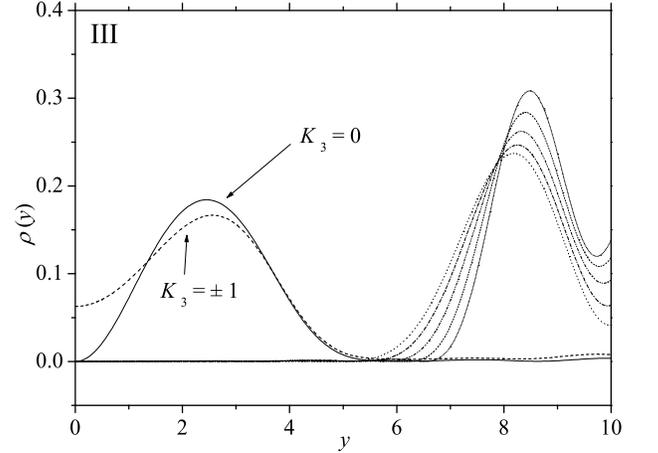}
\caption{\label{fdensity3} Fermion number density for the background $(m,n)=(1,3)$ skyrmion 
with the model parameters: $\alpha=0.11,~\tilde{\Lambda}_b=-0.35,~\tilde{\mu}=0.957648$. 
The case of the coupling constant $\tilde{M}=1.0$.
The states $K_3=0,\pm 1$ are localized on the 
brane core. Doubly degenerate $K_3=\pm 1$ states are more localized 
(they exhibit the non-zero value at the origin).}
\end{figure}

\begin{figure}
\includegraphics[height=6cm, width=8cm]{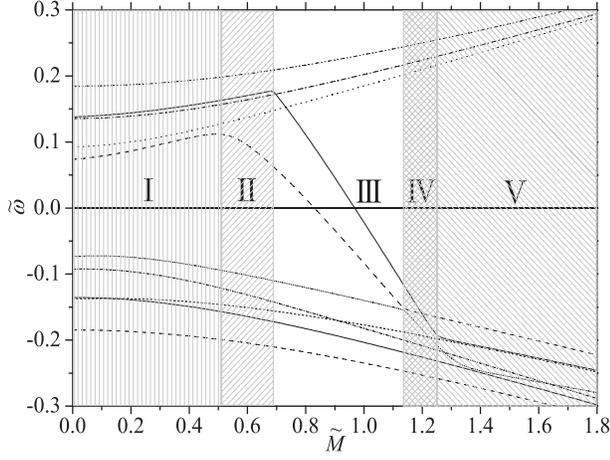}
\caption{\label{sflow3} Spectral flow of the fermion energy for the background $(m,n)=(1,3)$ 
skyrmion with the model parameters: $\alpha=0.11,~\tilde{\Lambda}_b=-0.35,~\tilde{\mu}=0.957648$. 
Two zero crossing spectra have the quantum number $K_3=0$ (straight line), and $K_3=\pm 1$ (dotted line),
respectively.}
\end{figure}

\subsection{\label{subsec:level32}The numerical method}
Instead of solving the second order differential equation (\ref{eigeneq2}), 
we study the eigenproblem (\ref{eigenproblem8}) directly.
We employ the method which was originally proposed by Kahana-Ripka~\cite{Kahana:1984be} 
for solving the Dirac equation with non-linear chiral background.
According to the Rayleigh-Ritz variational method \cite{bransden}, the upper 
bound of the spectrum can be obtained from the secular equation; 
\begin{eqnarray}
	\rm{det}\left(\bf{A}-\epsilon \bf{B}\right) = 0 
	\label{secular_equation}
\end{eqnarray}
with
\begin{eqnarray*}
     A_{ij}=
	\int d^{3}x \varphi_i^\dagger H
	\varphi_j,
	~~B_{ij}=
	\int d^{3}x \varphi_i^\dagger \varphi_j
\end{eqnarray*} 
where $\{\varphi_i\}~(i=1,\cdots,N)$ is some complete set of the plane-wave spinor basis. 
For $N \rightarrow \infty$, the spectrum $\epsilon$ becomes exact. 
Eq.(\ref{secular_equation}) can be solved numerically. 
For simplicity, we construct a plane-wave basis in large circular box with radius $D$ 
as a set of eigenstates of the flat, unperturbed ($B=1,B'=0,C=r,f=\pi$) Hamiltonian
\begin{eqnarray}
H_0=-\gamma^{\hat{6}}\gamma^{\hat{4}}\partial_4-\gamma^{\hat{6}}\gamma^{\hat{5}}\partial_5
-\gamma^{\hat{6}}\tilde{M}\tau^3\,.
\end{eqnarray}
 The solutions are 
\begin{eqnarray}
&&
\phi^{\rm up}_i
~=
M_{k_i}\left( 
\begin{array}{c}
\omega^{+}_{\epsilon_{k_i}}J_{p-1}(k_iy)e^{i(p-1)\theta} \\
\omega^{-}_{\epsilon_{k_i}}J_{p}(k_iy)e^{ip\theta}
\end{array}
\right)
\otimes
\left(
\begin{array}{c}
1 \\
0
\end{array}
\right) \nonumber \\
&&
\phi^{\rm down}_i
=
N_{l_i}\left( 
\begin{array}{c}
\omega^{-}_{\epsilon_{l_i}}J_{q}(l_iy)e^{iq\theta} \\
\omega^{+}_{\epsilon_{l_i}}J_{q+1}(l_iy)e^{i(q+1)\theta}
\end{array}
\right)
\otimes
\left(
\begin{array}{c}
0 \\
1
\end{array}
\right) \nonumber \\ 
\label{kahana_ripka}
\end{eqnarray}
with
\begin{eqnarray}
	M_{k_i}=\biggl[\frac{2\pi D^2\epsilon_{k_i}}{\epsilon_{k_i}+\tilde{M}}
	\Bigl(J_{p-1}(k_iD)\Bigr)^2\biggr]^{-1/2} \nonumber \\
	N_{l_i}=\biggl[\frac{2\pi D^2\epsilon_{l_i}}{\epsilon_{l_i}+\tilde{M}}
	\Bigl(J_{q+1}(l_iD)\Bigr)^2\biggr]^{-1/2}\,. \nonumber 
\end{eqnarray}
Here $\omega^{+}_{\epsilon_k<0},\omega^{-}_{\epsilon_k>0}=1, 
\omega^{-}_{\epsilon_k<0},\omega^{+}_{\epsilon_k>0}=-{\rm sgn}(\epsilon_k)k/(\epsilon_k+\tilde{M})$
and $\epsilon_k=\pm \sqrt{k^2+\tilde{M}^2}$.
The momenta $k_i,l_i~(i=1,\cdots,m_{\rm max})$ are discretized by the boundary conditions
\begin{eqnarray}
J_{p}(k_iD)=0,~~J_{q}(l_iD)=0
\end{eqnarray}
where $p:=K_3+\dfrac{1}{2}-\dfrac{n}{2}, q:=K_3-\dfrac{1}{2}+\dfrac{n}{2}$.
The orthogonality of the basis is then satisfied by  
\begin{eqnarray}
&&\int^D_0 dr r J_\nu(k_i r)J_\nu(k_j r)
=\int^D_0 dr r J_{\nu\pm 1}(k_i r)J_{\nu\pm 1}(k_j r)  \nonumber \\
&&=\delta_{ij}\frac{D^2}{2}  [J_{\nu\pm 1}(k_i D)]^2,~~
\nu:=K_3\pm\frac{1}{2}\mp\frac{n}{2}\,.
\label{orthogonality}
\end{eqnarray}
Expanding the eigenstates of Eq.(\ref{eigenproblem8}) in terms of the plane-wave basis, 
the eigenproblem reduces to the symmetric matrix diagonalization problem.  
A special care is taken for the estimation of the matrix element of the hamiltonian 
(\ref{hamiltonian8}). In order to hold the Hermiticity of the matrix, 
the following differential rule is imposed
\begin{eqnarray}
\langle \psi |\overleftrightarrow{\partial_y}|\phi\rangle =\int dyd\theta \tilde{C}(y)
\frac{1}{2}\Bigl[\psi^{\dagger}\partial_y\phi-(\partial_y\psi^{\dagger}) \phi\Bigr]\,.
\end{eqnarray}

\begin{figure}
\includegraphics[height=7cm, width=9cm]{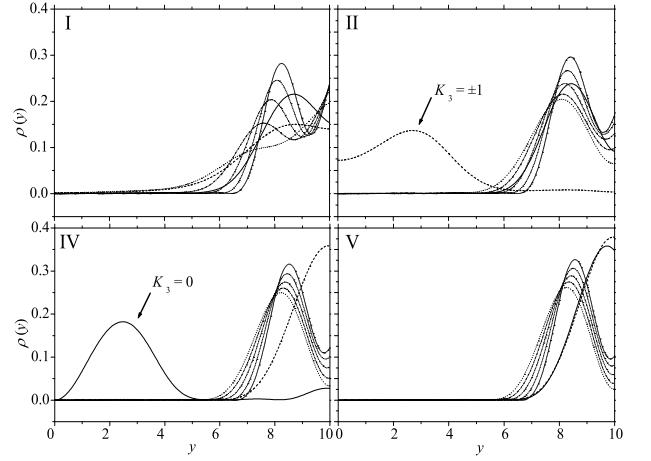}
\caption{\label{fdensity31} Fermion number densities for the background $(m,n)=(1,3)$ skyrmion 
about four regions depicted in Fig,\ref{sflow3}.}
\end{figure}

\begin{figure*}
\includegraphics[height=5cm, width=6cm]{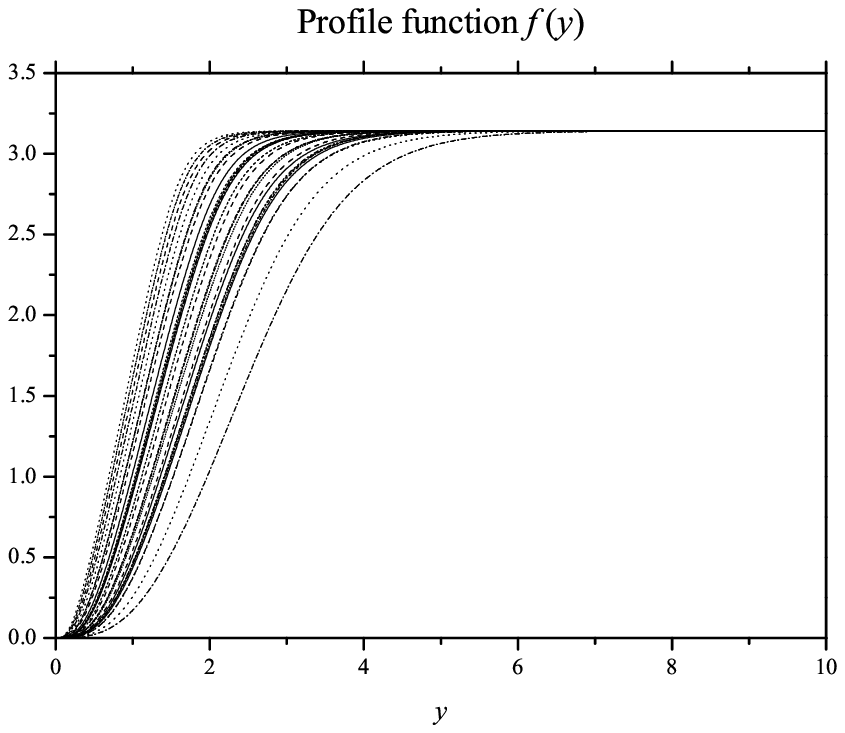}\hspace{-0.5cm}
\includegraphics[height=5cm, width=6cm]{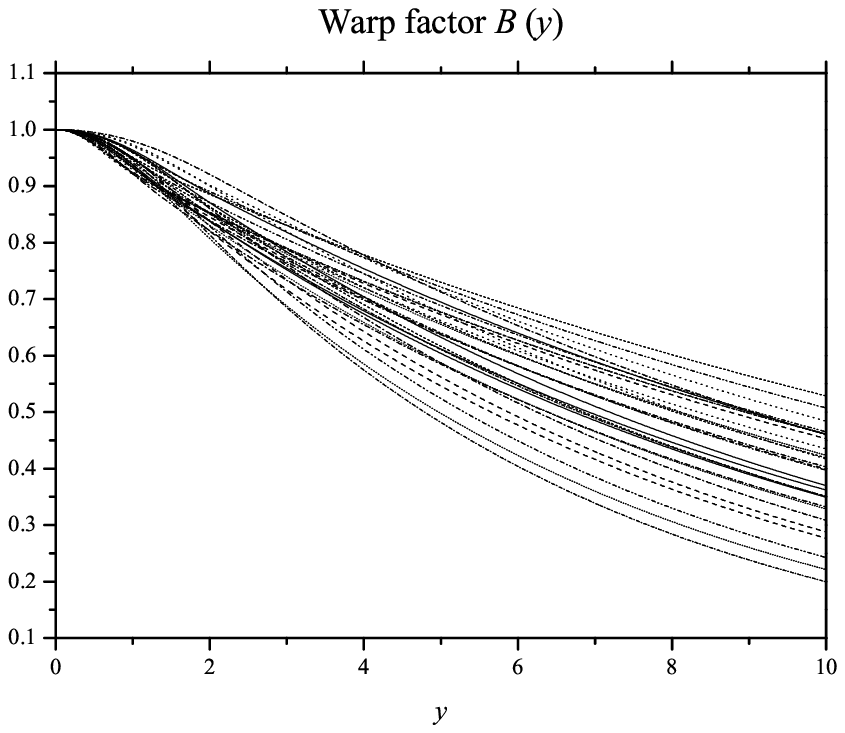}\hspace{-0.5cm}
\includegraphics[height=5cm, width=6cm]{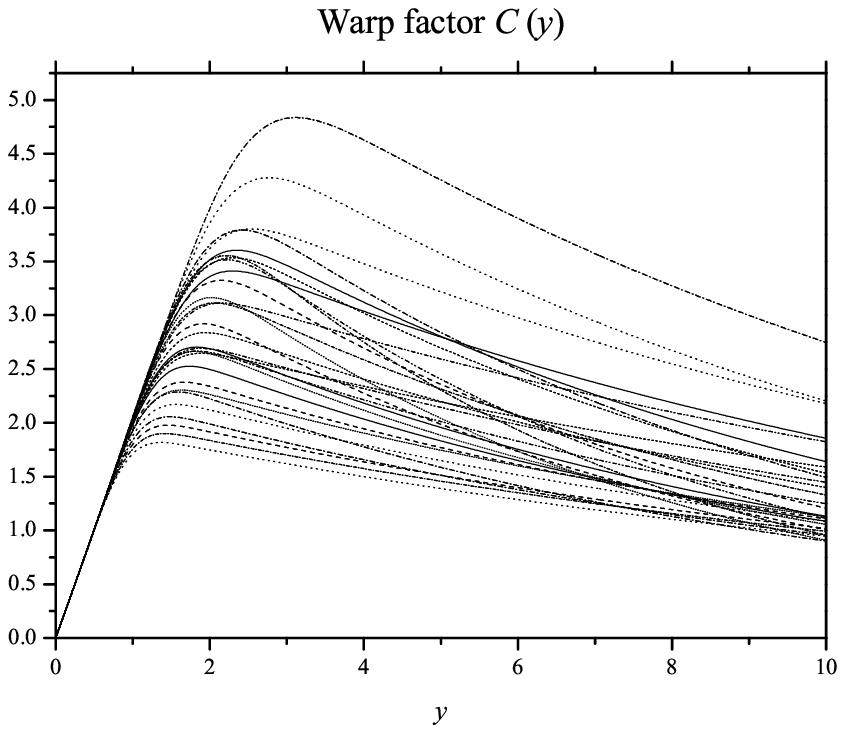}
\caption{\label{profilefb} The brane solutions with large parameter space are plotted.}
\end{figure*}

Size of the radius $D$ is chosen so as to wrap the whole branes. 
Apparently the typical value $D=10.0$ is sufficient because  
one easily observe that the brane profile functions, the stress 
energy tensors, and the curvature invariants are flat at 
$y>8$~(see Figs.[\ref{profilef},\ref{stensor},\ref{curvature}]). 

Fig.\ref{fdensity1} shows the densities $\rho(y)$ for the background skyrmion of $(m,n)=(1,1)$.
We display a tower of the massive modes together with the ground state. 
As one easily observe that only the lowest mode is peaked on the brane while all other modes
escape from the core; therefore the massive modes cannot be observed on the brane core. 
In Fig.\ref{sflow1}, we plot the fermion spectra as varying the coupling constant $\tilde{M}$. 
Only the lowest isolated mode decays from positive continuum to the negative. 
We will confirm that the spectrum corresponds to the brane localized mode. 

Libanov,Troitsky have discussed relation between the topological charge and the 
fermionic generation \cite{Libanov:2000uf}. 4-dimensional fermions appear as zero-modes trapped in the 
core of the global vortex with winding number three. 
We shall investigate this speculation with our solution for $(m,n)=(1,3)$. 
The results of the expansion at the origin (\ref{expf})-(\ref{expc}) are 
\begin{eqnarray}
{\cal B}=-\frac{\alpha}{4}\bigl(\tilde{\Lambda}_b+2\tilde{\mu}),~~
{\cal C}=\frac{\alpha}{12}\bigl(\tilde{\Lambda}_b+2\tilde{\mu})
\end{eqnarray}
and $f^{(3)}(0)$ is the shooting parameter. 
The typical example of solution of the brane is shown in Fig.\ref{profilef3}.
As is expected, we observe three localized solutions (Fig.\ref{fdensity3}) 
which can be regarded as the generations of the fermions.  
A main difference from the previous studies is that we use non-linear type of soliton solutions 
for constructing the branes. They have richer information of the topology and 
the spectra exhibit doubly degenerate ground states and a higher excited state. On the other hand, 
the linear type of solitons like Abelian vortex have the spectra with double degeneracy only. 
The mass spectra of the fermions in our universe comprises doubly degenerate plus 
single level with very high energy; it suggests the underlying topology that we employs.

\subsection{\label{subsec:level33}Fermion level crossing picture}
In Fig.{\ref{sflow1}}, we show that only one isolated mode dives from positive energy to 
negative. This behavior is called as spectral flow or level crossing picture~\cite{Kahana:1984be}. 
Spectral flow is defined as the number of eigenvalues of 
Dirac Hamiltonian that cross zero from below minus the number of eigenvalues that cross 
zero from above for varying the properties of the background fields. According to the 
Index theorem a nonzero topological charge implies the zero modes of the 
Dirac operator \cite{Atiyah:1980jh}. The number of flow coincides with the topological 
charge and the zero modes 
are emerged when they cross the zero. The level crossing picture extensively studied in the 
Dirac equation with non-linear chiral background~\cite{Kahana:1984be,Sawado:2004pm}, 
with the Higgs field of the Abelian-Higgs model~\cite{Bezrukov:2005rw,Burnier:2006za} and 
with the non-trivial gauge fields ($e.g.$,instanton,meron)~\cite{Christ:1979zm,Nielsen:1983rb,Kiskis:1978tb}.
In a electroweak theory, one level crossing of the fermion in the background of the spphaleron barrier 
is observed~\cite{Kunz:1993ir}.
Sometimes the mechanism can be thought of as a quantum mechanical description of fermion 
creation/annihilation. Interestingly, we can observe the mixing of some levels, $i.e.$,
the energy levels cannot cross and the excited particle changes with the light one 
(see Fig.\ref{sflow1}). 
This behavior indicates that these fermions interact with each other via some potential. 
This has been thoroughly discussed in Ref.\cite{Burnier:2006za} with somewhat different models.
In this subsection, we shall examine the level crossing behavior of the 
fermion in the brane-skyrmion background.

\begin{figure}
\includegraphics[height=7.0cm, width=9cm]{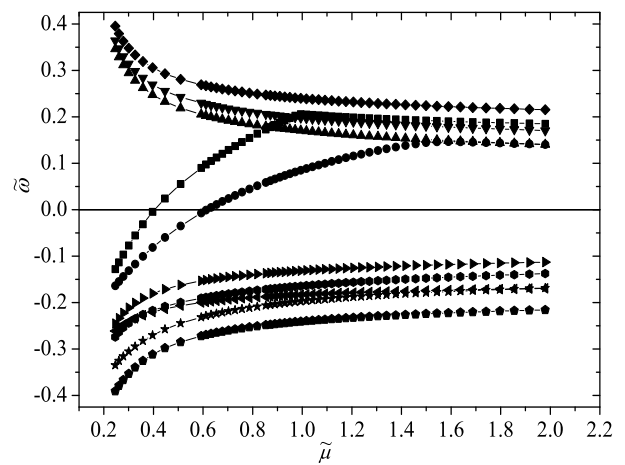}
\caption{\label{sflow31} Spectral flow of the fermion energy of $(m,n)=(1,3)$ for 
$\tilde{M}=0.72, \tilde{\Lambda}_b=-0.2$ and the different values of $\tilde{\mu}$ ($\alpha$
is the function of $\tilde{\mu},i.e., \alpha(\tilde{\mu})$).
Every dot corresponds to the different background brane solutions. Especially, 
the bullet ($\bullet$) is the lowest, doubly degenerate states and the square ($\blacksquare$) is the single
mode, respectively.}
\end{figure}
\begin{figure}
\includegraphics[height=7.0cm, width=9cm]{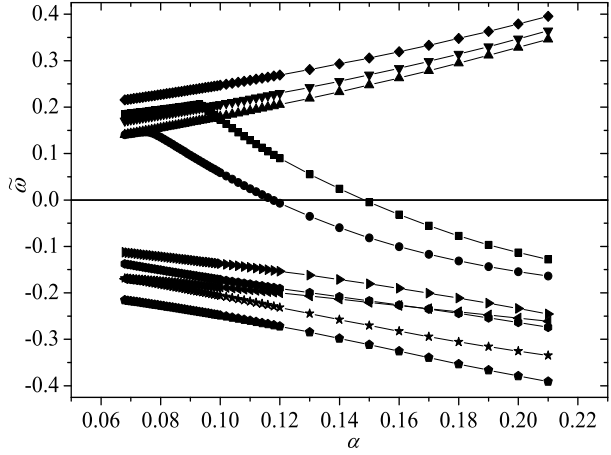}
\caption{\label{sflow32} Spectral flow of the fermion energy of $(m,n)=(1,3)$ for 
$\tilde{M}=0.72, \tilde{\Lambda}_b=-0.2$ and the different values of $\alpha$. 
The bullet ($\bullet$) is the lowest, doubly degenerate states and the square 
($\blacksquare$) is the single mode.}
\end{figure}

\begin{figure}
\includegraphics[height=7.0cm, width=9cm]{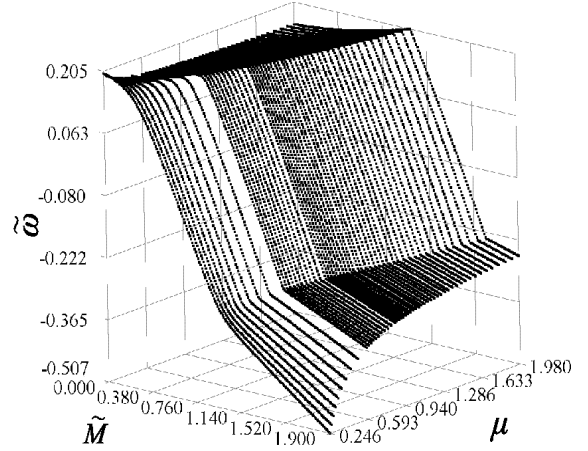}
\caption{\label{sflowb} Spectral flow ``cascade'' of one of the zero crossing mode 
in $(m,n)=(1,3)$, for fixed $\tilde{\Lambda}_b=-0.2$.
}
\end{figure}

Studying the spectral flow argument, the authors always investigate evolution of the 
spectrum by one parameter, $e.g.$, time (which characterize the size or the strength 
of the background field). Our model, however, has many parameters which define the 
basic property of the branes. Therefore, we explore the fermion level behavior 
for the brane solutions with large parameter space.
Fig.\ref{sflow3} shows the result of the spectral flow of the solutions with $(m,n)=(1,3)$ 
as a function of the coupling constant $\tilde{M}$. 
Fig.\ref{fdensity31} is the corresponding fermion densities. 
One easily observes that the density exhibits various localization behaviors in every domain.
In order to examine more thorough parameter dependence of the spectral flow, first 
we prepare several varieties of the brane solutions with large parameter space
 (see Fig.\ref{profilefb}). 
Once the brane solutions are obtained, the model parameters $(\alpha,\tilde{\mu},\tilde{\Lambda}_b)$ 
are no longer arbitrary, $i.e.$, a constraint $h(\alpha,\tilde{\mu},\tilde{\Lambda}_b)=0$ exists. 
So if we fix $\tilde{\Lambda}_b,\tilde{M}$, either $\alpha$ or $\tilde{\mu}$ is 
identified as a evolution parameter. 
We display in Figs.[\ref{sflow31},\ref{sflow32}], the spectral flow corresponding to these parameters. 
Fig.\ref{sflowb} shows more general spectral flow ``cascade'',$i.e.$, the flow as  
functions of $(\tilde{\mu},\tilde{M})$.

\subsection{\label{subsec:level34}Mass splitting of the generations}
Because of the time-reversal symmetry, the spectra contain
doubly degenerate states and, the first two generations of the 
fundamental fermions should be observed as the degenerate states within our framework. 
Some effects can split the degenerate states.
For example, in order to manifest the symmetry breaking, we introduce 
the explicit mass term for the fermions of the form
\begin{eqnarray}
H_m:=B\gamma^{\hat 6} {\hat m},~~
{\hat m}:=
\left( 
\begin{array}{cc}
m_1 & 0 \\
0  & m_2
\end{array}
\right)
\end{eqnarray}
$H_m$ commutes with the grandspin operator and then the eigenstate of the hamiltonian
with this term still are labeled by $K_3$. On the other hand, it
breaks the time-reversal symmetry, thus, this additional term successfully splits the degeneracy. 
Without loss of generality, we can set $m_1=0,m_2=\Delta m$, which is treated as a free
parameter. 

Another trial is based on a fine structure of the background solitons. 
In Ref.\cite{Hen:2007in}, the authors extensively studied the structure of the multisolitons with 
charge $1\le Q\le 5$ in the baby Skyrme model. They use one parameter family of the potentials 
$U=\mu^2(1-\phi_3)^s$. For the $Q=3$, the soliton has spherical symmetry below some critical 
value $s$ (our calculation is the case of $s=1$). 
The symmetry is broken above the critical value, and the solution exhibits only $\mathbb{Z}_2$ symmetry. 
So, we expect that our branes slightly deform and the degenerate spectra split. 
 
For the $3+1$ Skyrme model, construction of the ansatz for the multisolitons with spherical coordinates 
$(r,\theta,\phi)$ called {\it Rational map ansatz} 
has been proposed by Houghton, Manton and Sutcliffe \cite{Houghton:1997kg}. 
Although the solitons exhibit the complex
platonic symmetries, they become tractable by using this ansatz. 
Coupling of the fermions to the Rational map 
skyrmions was already studied in Refs.\cite{Sawado:2002zm,Sawado:2004pm}.
In the baby Skyrme model, however, no analytical form of ansatz for such 
non-continuous symmetry exists, then we treat the effects as a small perturbation.  

In general, for the solution with non spherical symmetry, the profile function is modified $f(r)\to F(r,\theta)$.
In the case of $\mathbb{Z}_2$, if the deformation is small, $F(r,\theta)$ can be expanded
\begin{eqnarray}
F(r,\theta)\sim f(r)+\frac{\tilde{\epsilon}}{2}f_1(r)(e^{2i\theta}+e^{-2i\theta}) 
\end{eqnarray}
where the first term of the right hand side $f(r)$ corresponds to the profile function with spherical symmetry. 
One can easily confirm that the solution has $\mathbb{Z}_2$ symmetry, $i.e.$, $F(r,\theta+\pi)=F(r,\theta)$.
Substituting into Eq.(\ref{hedgehog}), the term 
$\vec{\tau}\cdot\vec{\phi}\to \vec{\tau}\cdot\vec{\phi}+V_\epsilon$
thus hamiltonian (\ref{hamiltonian8}) can be written as 
\begin{eqnarray}
H_2\to \bar{H}_2(\epsilon)\equiv H_2+B\gamma^{\tilde{6}}\tilde{M}V_\epsilon
\end{eqnarray}
where the potential $V_\epsilon$ is defined as 
\begin{eqnarray}
&&V_\epsilon(r,\theta;\epsilon) := \nonumber \\
&&\epsilon\left( 
\begin{array}{cc}
-\sin f(r)(e^{2i\theta}+e^{-2i\theta}) & \cos f(r)(e^{-i\theta}+e^{-5i\theta}) \\
\cos f(r)(e^{i\theta}+e^{5i\theta})  & \sin f(r)(e^{2i\theta}+e^{-2i\theta})
\end{array}
\right)\,.
\nonumber \\
\end{eqnarray}
for the charge three ($n=3$). 
Since we have no detailed information about the deformation, 
we suppose $\epsilon:=\frac{\tilde{\epsilon}}{2}f_1(r)$ as a constant parameter
describing measure of the deformation.
The potential breaks both the time-reversal and the grand spin symmetry, 
then the coupling between states with different grandspin $K_3$ occur. 
Mixing of states $(K_3,K'_3)$ with large $\Delta K_3:=|K_3-K'_3|$ 
can be negligible for small deformation. 
Here we compute the eigenproblem $\bar{H}_2(\epsilon) u=\bar{w} u$, 
taking into account only the coupling with $\Delta K_3\leqq 2$.

\begin{figure}[t]
\includegraphics[height=7cm, width=9cm]{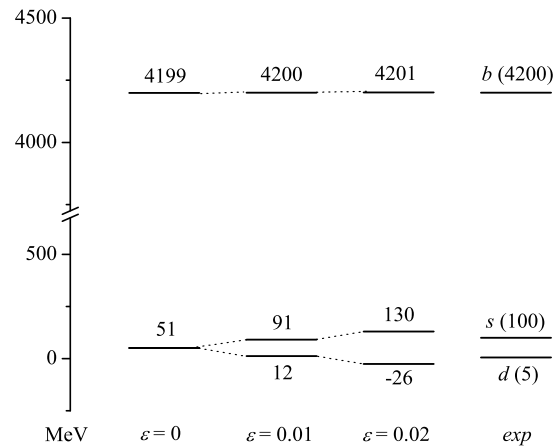}
\caption{\label{masssplit1} Mass splitting of the down series of quarks $(d,s,b)$ 
in terms of effect of the deformation of the background baby skyrmions as  
function of the parameter $\epsilon$ which determines the 
strength of the deformation (in MeV). 
The experimental values are quoted from Ref.\cite{Amsler:2008zz}.}
\end{figure}

\begin{figure}[t]
\includegraphics[height=7cm, width=9cm]{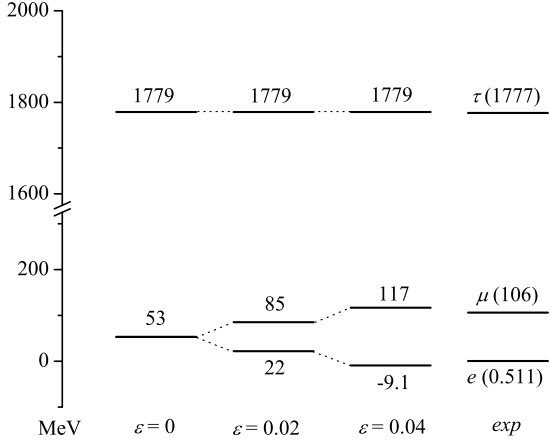}
\caption{\label{masssplit2} Mass splitting of the leptons $(e,\mu,\tau)$ 
in terms of effect of the deformation of the background baby skyrmions as 
function of the parameter $\epsilon$ which determines the 
strength of the deformation (in MeV).
The experimental values are also from Ref.\cite{Amsler:2008zz}.}
\end{figure}

At present status of our model, we admit that it is a toy model 
for understanding the level structure of the realistic standard model 
fermions because we have no explicit realistic charges. 
We are able to fit just one quark/lepton sector in terms of a set of model parameters. 
First we evaluate the mass of the ``down sector'' of quarks $(d,s,b)$.
Fig.\ref{masssplit1} shows the spectra. We employ a brane solution with the parameters
$\alpha=0.08, \tilde{\Lambda}_b=-0.21, \tilde{\mu}=1.4029$, and the coupling constant $\tilde{M}=0.94$.
For obtaining the dimensionful values, we tentatively choose the parameters of the Skyrme model, 
$Fe=27800$ MeV. We obtain masses of the down, the strange, 
and the bottom quark $m_d=5.1$ MeV, $m_s=98$ MeV, $m_b=4200$ MeV, for $\epsilon=0.012$. 
For the lepton sector, we have to employ a different solution which is characterized by the 
set of parameters $\alpha=0.07,\tilde{\Lambda}_b=-0.21, \tilde{\mu}=1.8838$ and $\tilde{M}=1.03$. 
The Skyrme model parameter $Fe=10506$ MeV and when the deformation parameter is $\epsilon=0.03383$, 
we obtain the masses $m_e=0.5$ MeV, $m_\mu=107$ MeV, $m_\tau=1779$ MeV (Fig.\ref{masssplit2}).
These are in quite good agreement with the corresponding experimental observations. 

Here, it is worth mentioning the physical implications of our parameter choice. 
In Sec.\ref{sec:2}, we have computed the brane solutions with wide range of the parameters
$(\alpha,\tilde{\Lambda}_b,\tilde{\mu})$, which are physically equivalent. 
The coupling constant $\tilde{M}$ has been chosen in order that the lowest energy levels cross zero.
The splitting of the first two generations and the third can be adjusted in terms 
of the choice of $(\alpha,\tilde{\Lambda}_b,\tilde{\mu})$.
Essentially, the deformation parameter $\epsilon$ is not a free parameter;
it should be determined uniquely in terms of a consistent calculation of the Einstein-Skyrme systems. 
Thus, one can say that the splitting of the first two generations is also 
controlled by the brane parameter choice. 
Finally, the skyrmion parameter $Fe\sim 10^4$ MeV guarantees  
the size of the branes as $r\sim 10^{-1}$ fm, which is consistent with our observation;
it is sufficiently small so as not to observe any evidence of the extra dimension.  
 
\section{Conclusion}
In this article, we have proposed new brane solutions in 6-dimensional space-time 
and have discussed about localization of the fermions on them. 
The brane have constructed by a baby-skyrmion, a generalization of a O(3) non-linear $\sigma$ model, 
and have realized the regular warped compactification in 6-dimensional anti-de Sitter space-time. 
The metric is non-factorizable with the warp factors either exponentially diverging 
or decaying in static solutions. But only exponentially decaying warp factors 
can allow gravity localization near the brane in the sense that the 6-dimensional Planck 
mass takes a finite value. 
Such solutions could be obtained numerically if we impose suitable boundary conditions 
at the distance far from the brane and integrate in backward. On the other hand,  
the forward integration method is unstable for any boundary conditions at the origin.

After the solutions of the skyrmion branes were successfully found, we have studied 
the fermion localization on them. 
The Dirac equation in 6-dimensional curved space-time has been constructed in terms of 
introducing the vielbein and 6-dimensional generalization of the gamma matrices. 
To treat the massive modes, we have used the standard representation of the gamma 
matrices. In order to solve the eigenproblem, we have introduced the plane wave basis in
a large circular box with radius $D$.  Studying identification of the charge 
of the background skyrmions and the generation number of the standard model fermions
in our universe, we have investigated the skrymions of the topological charge $Q=3$. 
This conjecture has been confirmed through the discussion of the fermion spectral flow. 
It embodies the localized modes where the number equals to the background topological charge.
As a result, we have found localized fermion modes on the skyrmion branes corresponding to their
topological charge. For $(m,n)=(1,1)$, we have observed the solutions with some parameter ranges, 
which means the existence of the massive modes as well as the massless one.
For $(m,n)=(1,3)$, three solutions localized on the brane have been found. 
They comprise doubly degenerate lowest modes of plus single excited state.
This level structure well describes the experimental measurements of fermion masses. 
On the other hand, in the Higgs mediated models, the spectra exhibit only double 
degeneracy.

As was found in Ref.\cite{Hen:2007in}, the minimal energy configuration of the baby skyrmion with charge three
has no spherical symmetry, rather, it exhibits $\mathbb{Z}_2$ symmetry. The small shape deformation
has an effect on the lowest degenerate state and splits them. We have successfully 
obtained the tower of the ``down''sector of quark generation within one parameter family of the 
deformation. We have treated the deformation of the skyrmion in a perturbative way and
have neglected the effect for the gravity. In order to address the deformation of the brane solutions 
from spherical geometry, we need to examine full simulation of the Einstein-Skyrme system 
without imposing any symmetrical ansatz for the geometry. 
We believe that the solution will clarify the origin of the level structure of the standard model fermions. 

In Ref.\cite{Ringeval:2001cq}, the authors had an elaborate analysis for the massive modes in a 
5-dimensional anti-de Sitter space-time.
They set a domain that is defined by the effective potential, and then solved the Dirac equation with
inside and out. By imposing the boundary conditions at the surface, they obtained several 
massive modes. In our case, we also introduce a domain particularly for the numerical reason. 
We have confirmed that even if the radius of the domain goes to its infinity, 
there is no notable effect to the zero mode. 

Our model is a simple toy model for understanding the level structure of the realistic standard model
fermions. 
We observed generation of masses of the down sector $(d,s,b)$ or the lepton sector $(e,\mu,\tau)$ 
with good agreement of experimental observations. 
Perhaps most serious drawback of our approach is that we have much number of free parameters 
and need the independent parameter sets to fit each of the quark/lepton sectors. 
However, this may be justified if one considers that the quarks and leptons are localized on 
different branes but share the 4-dimensional spacetime. 
Another defect of our model is of course lack of the explicit spin and charge of the fermions.
Taking into account the effects of the breaking of the flavor symmetry and assigning the realistic 
charges for all quark/lepton sectors to our solutions will be a great advance toward 
full understanding of the generation mechanism of our universe. 

Our choice of the representation in this article might need more thorough consideration. 
Although there are several advantages to treat the massive modes, 
mechanism of localization of left-chiral fermions on the brane is not clear. 
Estimation of the asymmetry of the chirality on the brane is undoubtedly important subject, 
at least, about the lowest zero crossing level. 
The result implementing this will be reported in the forthcoming paper. 

\section{Acknowledgement}
We appreciate to Noriko Shiiki for giving us many useful advices and comments. 

\appendix
\section{The effective potential}\label{ap:potential}
The explicit form of $V_{ij}$ In Eq.(\ref{eigeneq2}) are given by
\begin{widetext}
\begin{eqnarray}
	V_{11}&=&-\beta^{-1}+\biggl(\delta^{l+1}\beta^{-1}-\beta\frac{n}{\tilde{C}}\tilde{M}^2\sin^2{f}
	+\beta G_+G'_--\frac{1}{2}\beta \tilde{M}^2 \sin 2ff'\biggr)\frac{l}{\tilde{C}}-\frac{l\tilde{C}'}{\tilde{C}^2} \nonumber \\
		&&-\frac{{\cal R}'-{\cal A}_1'}{2}
		-\left(\frac{{\cal R}-{\cal A}_1}{2}\right)\left(\frac{{\cal R}+{\cal A}_1}{2}\right)\\
	\label{V11}
	V_{12}&=&-\beta\tilde{M}\frac{l+n}{\tilde{C}}\sin f\left(G_+\cot ff'-G'_+-\frac{n}{\tilde{C}}G_+\right)-{\cal A}_2\frac{{\cal R}-\tilde{\cal A}_2}{2}\\
	V_{21}&=&-\beta\tilde{M}\frac{n}{\tilde{C}}\sin f\left(G_-\cot ff'-G'_-+G_-\frac{n}{\tilde{C}}\right)-\tilde{\cal A}_1\frac{{\cal R}-{\cal A}_1}{2}\\
	 \label{V12}
	V_{22}&=&-\beta^{-1}+\left(\delta^{l+1}\beta^{-1}+\beta\frac{n}{\tilde{C}}G_+G_-+\beta G_-G'_+
	-\frac{1}{2}\beta\tilde{M}^2\sin 2ff'\right)\frac{l+n}{\tilde{C}}-\frac{(l+n)\tilde{C}'}{\tilde{C}^2}\nonumber\\
		&&-\frac{{\cal R}'-\tilde{{\cal A}}_2'}{2}
		-\left(\frac{{\cal R}-\tilde{{\cal A}}_2}{2}\right)\left(\frac{{\cal R}+\tilde{{\cal A}}_2}{2}\right)
	\label{V22}
\end{eqnarray}
where
\begin{eqnarray}
&&{\cal A}_1:=\delta^1\beta^{-1}
+\beta\left(G_+ G'_-- \tilde{M}^2\sin^2 f\frac{n}{\tilde{C}}-\frac{1}{2}\tilde{M}^2\sin 2f f'\right)~~~~
\label{A1} \\ 
&&{\cal A}_2:=\beta\tilde{M}\sin f\left(\frac{n}{\tilde{C}}G_+-G_+\cot ff'+G'_+\right)
\label{A2} \\
&&\tilde{{\cal A}}_1:=\beta\tilde{M}\sin f\left(\frac{n}{\tilde{C}}G_--G_-\cot ff'+G'_-\right)
\label{AA1} \\ 
&&\tilde{{\cal A}}_2:=\delta^1\beta^{-1}
+\beta\left(G_- G'_++ \tilde{M}^2\sin^2 f\frac{n}{\tilde{C}}-\frac{1}{2}\tilde{M}^2\sin 2ff'\right).
\label{AA2}
\end{eqnarray}
\end{widetext}

\end{document}